\documentclass[11pt,a4paper]{article}

\usepackage{graphicx}

\usepackage{authblk}

\usepackage{cite}

\usepackage{hyperref}

\usepackage{geometry}
\usepackage{amsmath,amssymb}
\usepackage{caption}
\usepackage{subcaption}

\newcommand{\be}{\begin{equation}}

\newcommand{\ee}{\end{equation}}

\newcommand{\bea}{\begin{eqnarray}}

\newcommand{\eea}{\end{eqnarray}}

\newcommand{\ba}[1]{\begin{array}{#1}}

\newcommand{\ea}{\end{array}}

\newcommand{\eqrf}[1]{Eq.\ (\ref{#1})}

\title{A New f(R) Model in the Light of Local Gravity Test and Late-time Cosmology }

\date{}

\author[a,b]{Akhilesh Nautiyal\footnote{akhilesh.phy@mnit.ac.in}}
\author[b]{Sukanta Panda\footnote{sukanta@iiserb.ac.in}}
\author[b]{Avani Patel\footnote{avani@iiserb.ac.in}}

\affil[a]{Department of Physics, Malaviya National Institute of Technology Jaipur, Jaipur 302 017, India}
\affil[b]{Department of Physics, Indian Institute of Science Education and Research Bhopal, Bhopal 462 066, India}

\begin{document}

\maketitle

\begin{abstract}
  We propose a new model of $f(R)$  gravity containing Arctan function in the lagrangian. We show here that this model satisfies fifth force constraint unlike a similar model in Kruglov 2013. In addition to this, we carry out the fixed point analysis as well as comment on the existence of curvature singularity in this model. The cosmological evolution for this $f(R)$ gravity model is also analyzed in the Freidmann Robertson Walker(FRW) background. To understand observational significance of the model, cosmological parameters are obtained numerically and compared with those of Lambda cold dark matter ($\Lambda$CDM) model. We also scrutinize the model with supernova data. We apply Om diagnostic given by Sahni et al. 2008 to the model. Using this diagnostic, we detect the distinction between cosmic evolution caused by the $f(R)$ model and $\Lambda$CDM. We find best-fit parameter values of the model using Baryon Acoustic Oscillations data.
\end{abstract}

\section{Introduction}\label{intro}
Recent progress in the precise measurements of cosmic microwave background anisotropies and observations of type Ia supernovae and large scale structure strongly indicate that our universe, at present, is expanding acceleratingly\cite{Nobel2011}. An extra unknown component called dark energy is required to dominate over all other matter in recent times to make General Relativity (GR) enable to explain such accelerating expansion. These observations also suggest that the equation of state parameter of the unknown source should be negative. The simplest model which accommodates these features and is consistent with most of the observations is $\Lambda$CDM model, where $\Lambda$ cosmological constant. The required value of the cosmological constant to fit with observations is many orders of magnitude smaller than the value encountered from standard field theoretical calculations of vacuum energy \cite{Padilla:2015aaa}. This large discrepancy forces us to look for alternative explanations for dark energy. One of such alternatives is to replace the Ricci scalar, $R$, in Einstein-Hilbert action, by $f(R)$, a general function of $R$, which is known as $f(R)$ theories of gravity \cite{f(R)TheoriesReview,Sotiriou:2008rp}. If $f(R)$ theory can mimick $\Lambda$CDM at high redshift without using cosmological constant explicitly then the above discripancy can be circumvented. The $f(R)$ model studied here gives rise to effective cosmological constant in a sufficiently curved spacetime. This behaviour is a purely curvature induced effect and not related to vacuum energy. 

Since observations favor cosmological constant as a best fit dark energy candidate, natural possibility for $f(R)$ model is to construct one that can mimic as an effective cosmological constant today and at the same time it should be distinguishable from $\Lambda$CDM model at recent times. With this clue, many $f(R)$ models were proposed \cite{Hu:2007nk,Starobinsky:2007hu,Appleby:2007vb,Appleby:2009uf,Tsujikawa:2007xu,Zhang:2005vt,Cognola:2007zu,Miranda:2009rs,Linder:2009jz,Bamba:2010ws,Bamba:2010zz,Kruglov:2013qaa}  
which reduces to the $\Lambda$CDM model in large curvature limit i.e. $f(R)\rightarrow R-2\Lambda$ for $R>>\Lambda$ and tends to zero as $R \rightarrow 0.$  The dynamical system analysis of $f(R)$ theories are done in Ref.\cite{Amendola:2006we} wherein these theories are classified according to their fixed points. A new approach for dynamical system analysis of $f(R)$ models is established in \cite{Carloni:2015jla} and applied to few viable models in \cite{Carloni:2015jla,Kandhai:2015pyr}. Any modified gravity theory should be put to the test to inquire that it does not spoil the successes of GR at local scales like the solar system. In general, all $f(R)$ theories are effective scalar-tensor theories with a potential term. If the mass of the scalar field is order of present Hubble parameter then it is difficult to satisfy the local gravity constraints due to the long-range fifth force with a large coupling strength \cite{Will:2014xja,Olmo:2005hc,Olmo:2005zr}. There must exist some screening mechanism which screens the fifth force in high-density region.  

Many $f(R)$ models are also plagued by fatal curvature singularities of various types. The cosmological evolution happens at the minimum of the scalar field potential around which scalar field oscillates.  The point of diverging curvature exists near the minimum at a finite potential in many $f(R)$ models \cite{Appleby:2008tv,Frolov:2008uf,Dutta:2015nga,Dutta:2016ukw}. The finite-time singularity in modified gravity is described in \cite{Nojiri:2008fk,Bamba:2008ut,Dev:2008rx,Thongkool:2009js,Capozziello:2009hc,delaCruz-Dombriz:2015tye}. It is also realized that the curvature singularities can be eliminated by adding an $R^2$ term to the Lagrangian \cite{Appleby:2008tv,Appleby:2009uf}. The curvature singularity can also be seen in an astrophysical object \cite{Arbuzova:2010iu,Lee:2012dk,Reverberi:2012ew,Dutta:2015nga,Dutta:2016ukw}. The extensive investigations of singularities in relativistic stars are done in \cite{Kobayashi:2008tq,Upadhye:2009kt,Babichev:2009fi}. 

Though many modified gravity theories are becoming quite successful in explaining late-time acceleration, they suffer from the problem that their physical implications cannot be distinguished from each other. Many modified gravity models are also indistinguishable from $\Lambda$CDM. To solve this, we need an observation based, a model-independent test which can distinguish above models physically. A two-point diagnostic namely Om diagnostic to distinguish evolving dark energy models and $\Lambda$CDM model is given in \cite{Shafieloo:2012rs}. The Om diagnostic is originally proposed in \cite{Sahni:2008xx} as a null test of Dark Energy being a cosmological constant $\Lambda$. The brief discription of Om diagnostic is given in Sec.~\ref{sec:omh2}. This two-point diagnostic is slightly modified in \cite{Sahni:2014ooa} by multiplying by $h^2$ which we call $Omh^2$ diagnostic. An important property of the $Omh^2$ diagnostic is that it uses expansion history $H(z)$, whose value can be reconstructed from observations of luminosity distance $D_L$ via a single differentiation. This $Omh^2$ diagnostic can also provide a way to check the viabilities of modified gravity models and to distinguish them from each other as well as from $\Lambda$CDM. It is applied to Starobinsky model\cite{Starobinsky:2007hu} and Hu-Sawicki\cite{Hu:2007nk} model of $f(R)$ gravity in \cite{Jaime:2015afa}. In recent paper \cite{Nunes:2016drj}, observational constraints on $f(R)$ models are obtained from cosmological chronometer, supernova and baryon acoustic oscillations data.

In \cite{Dutta:2016ukw}, a detailed study of local gravity tests and fixed-point analysis is done for an Arctan model proposed in\cite{Kruglov:2013qaa}. It turns out that this simple Arctan model is ruled out by fifth-force constraint. In this work, we propose a new Arctan model of $f(R)$ gravity extending the model in\cite{Kruglov:2013qaa}. We check the viability of the model through local gravity test. In addition to this, we also carry out dynamical system analysis and find the fixed points for this model, following the line of \cite{Amendola:2006we}. Through the fixed point analysis, we show that our model and Hu-Sawicki model \cite{Hu:2007nk} belong to two different class of models. Cosmological evolution using this model is also presented by solving FRW equations numerically as described in \cite{Jaime:2012gc}. The distance modulus calculated in this model is fitted with the supernova data given by \cite{Amanullah:2010vv} for certain parameters. We investigate distinguishability of new $f(R)$ model from $\Lambda$CDM model using $Omh^2(z_i,z_j)$ diagnostic. We also find the best parameter values for the model by carrying out the $\chi^2$ test of $Omh^2$ using Baryon Acoustic Oscillations data. 

This paper progresses as follows. In the next section, we give a general idea of the modified Arctan model. The de Sitter points are obtained in Sec.~\ref{subsec:desitter} and Sec.~\ref{FPA} discusses the fixed points of the model and their stability. The fifth-force test and its result are explained in Sec.~\ref{sec:fifthforce}. The investigation of curvature singularity is carried out in Sec.~\ref{sec:cursingularity}. 
Sec.~\ref{sec:cosmoevol} presents cosmological evolution and behavior of density parameters and equation of state with the given modified Arctan model. The luminosity distance and distance modulus in this model along with the Union 2 compilation of supernova data is given in Sec.~\ref{sec:lddm}. The $Omh^2$ Diagnostic of the $f(R)$ model is carried out in Sec.~\ref{sec:omh2}. Finally, conclusions are discussed in Sec.~\ref{sec:conclusion}.

\section{A New $f(R)$ Model}\label{sec:newarctan}
\begin{figure}
 \centering
 \includegraphics[width=0.5\textwidth]{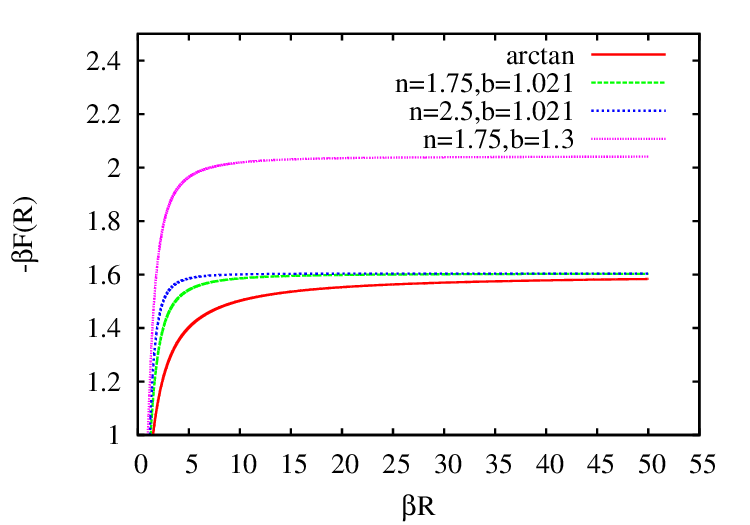}
 \caption{$-\beta F(R)$ Vs. $\beta R$. In "arctan" curve $n=1$ and $b=1.021$ are taken.}
 \label{fig:functionnewarctan}
\end{figure}

In $f(R)$ theory, Hilbert-Einstein action is replaced by a more general action
\begin{equation}
S=\int d^4x \sqrt{-g}\left[\frac{1}{2\kappa^2} f(R)+\mathfrak{L}_m\right],
\label{action}
\end{equation}
where $f(R)$ is an arbitrary function of Ricci scalar $R$ and $\mathfrak{L}_m$ is usual matter Lagrangian. It is more convenient to write the function $f(R)$ as $f(R)=R+F(R)$, where $F(R)$ is an arbitrary function of $R$. It can be easily seen that one can retain GR and $\Lambda$CDM for $F(R)=0$ and $F(R)=-2\Lambda$ respectively. Another benefit of writing $f(R)$ in this way is that effects of modification to Einstein gravity is more evident and its study becomes simpler. Here, we propose a new model of $f(R)$ theory, slightly modifying Arctan model given in \cite{Kruglov:2013qaa},
\begin{equation}
F(R)=-\frac{b}{\beta}Arctan(\beta R)^n.
\label{model}
\end{equation} 
Here, $b$, $\beta$ and $n$ are positive constants. $\beta$ has the inverse dimension of Ricci scalar $R$ and it is of the order of inverse of presently observed effective cosmological constant. Note that our model reduces to the Arctan model proposed in \cite{Kruglov:2013qaa} for $n=1$. Whenever we write "Arctan model" throughout this paper we mean the model in \cite{Kruglov:2013qaa}. In Fig.~\ref{fig:functionnewarctan}, the function $-\beta F(R)$ is plotted w.r.t. $\beta R$ for different values of $n$.
One can see from the Fig.~\ref{fig:functionnewarctan} that the variations in $\beta F(R)$ occurs only between $\beta R\sim 1-5.$ Beyond $\beta R\sim 5$, the value of $\beta F(R)$ becomes nearly constant. This behaviour tells us that our model given in \eqrf{model} mimics as $\Lambda$CDM at high curvature i.e. at $R>>1/\beta$ and $f(R)$ becomes zero at $R=0$. Also, notice that the function $F(R)$ increases and becomes constant more rapidly as $n$ increases,  which suggests that the value of parameter $n$ should be small enough so that the model can explain correct cosmic history. While keeping $n$ constant, the shape of the curve remains same, but its magnitude increases with increasing $b$.  

\subsection{Constant Curvature Solution}
\label{subsec:desitter}
\begin{figure*}
  \centering
   $
   \begin{array}{c c}
   \includegraphics[width=0.48\textwidth]{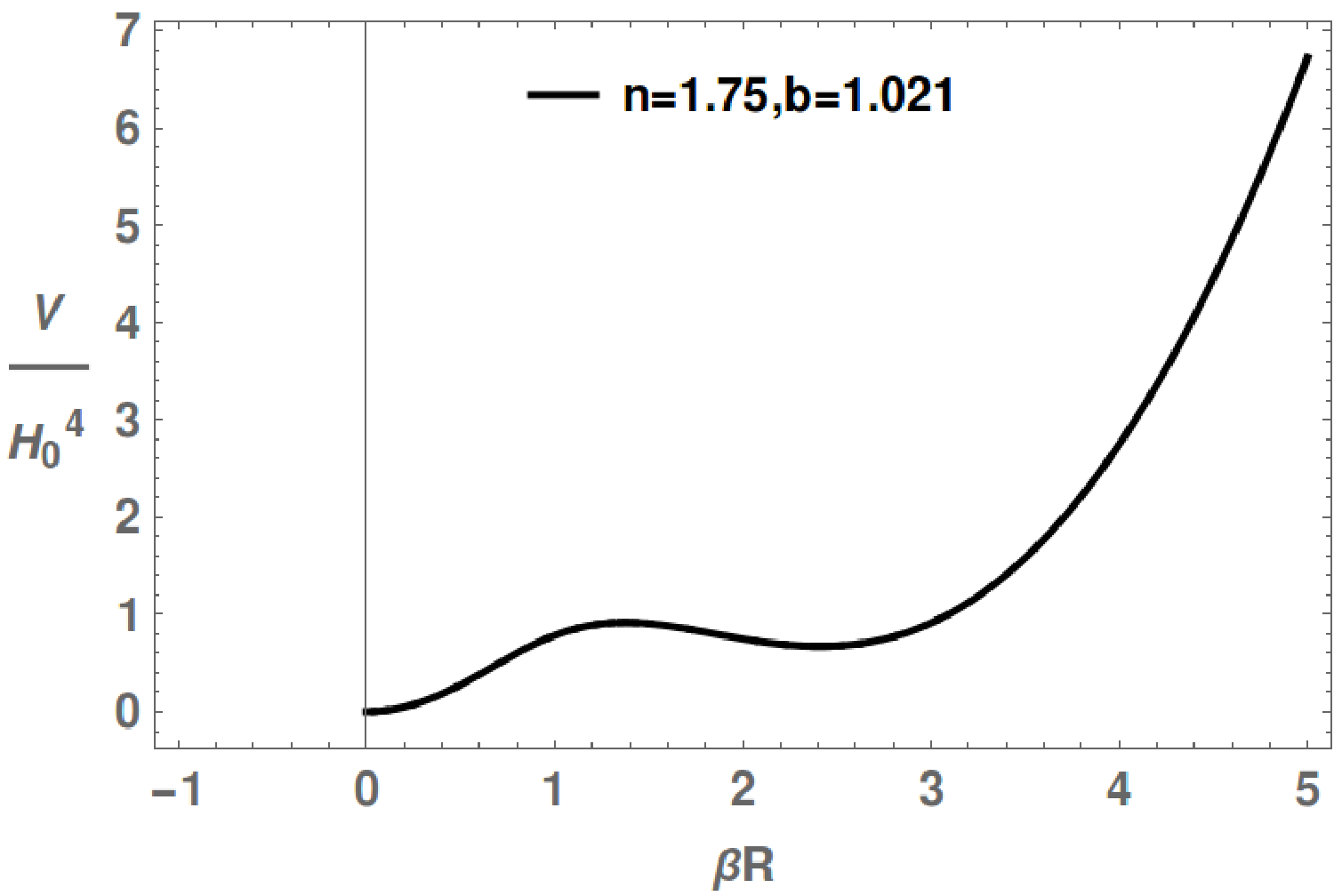} & 
   \includegraphics[width=0.48\textwidth]{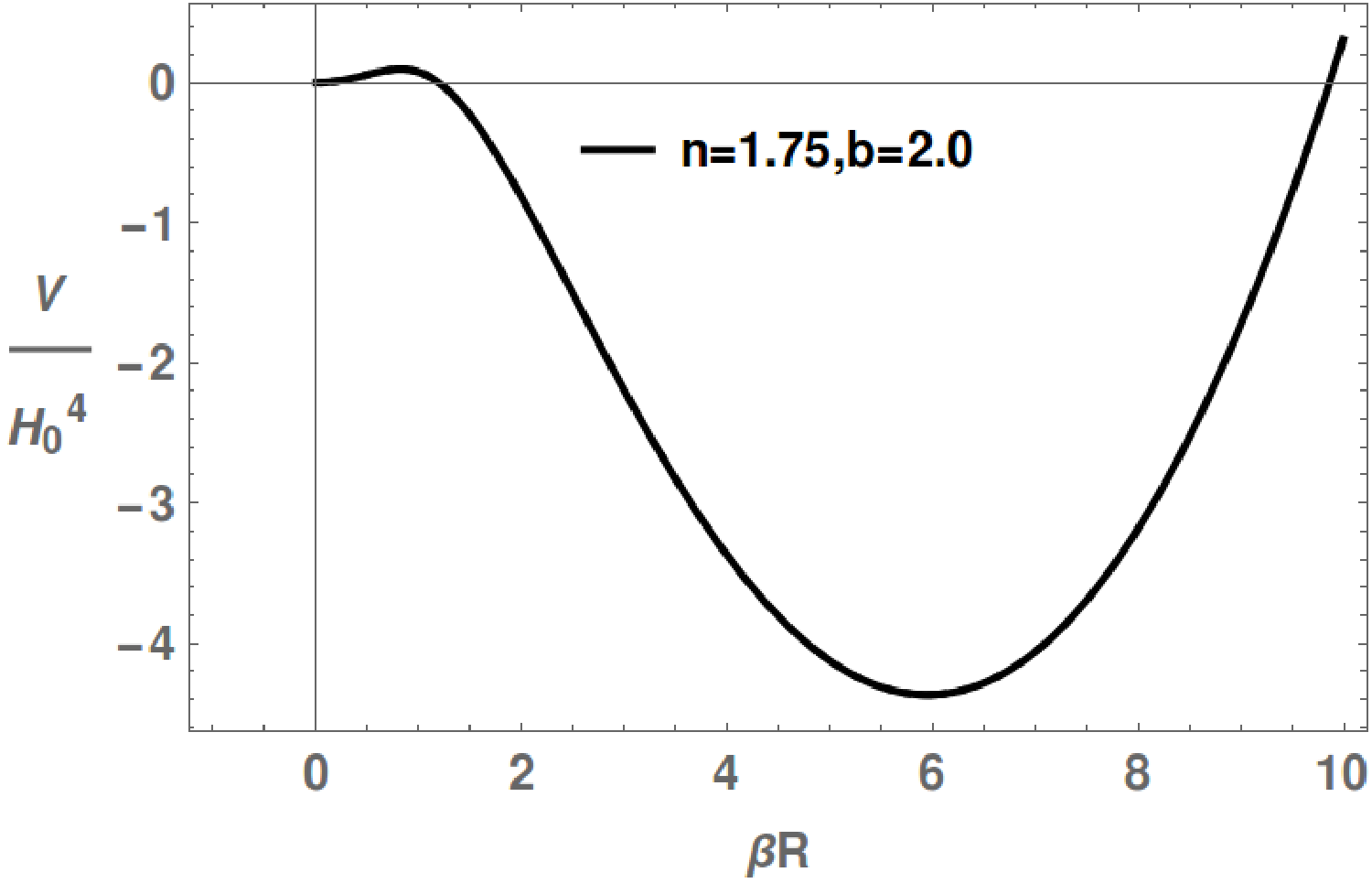}
   \end{array}
   $ 
  \caption{$V(R)/H_0^4$ vs $\beta R$ for the model (\ref{model}) and the potential $V(R)$  is defined by
\eqrf{modelpot}}
  \label{fig:modelpot}
\end{figure*}

By varying the action written in \eqrf{action} w.r.t. $g_{\mu\nu}$ we obtain the field equation given as,

\be
f_{,R} R_{ab} -\frac12 fg_{ab}-\left(\nabla_a\nabla_b-g_{ab}\Box\right)f_{,R} = \kappa^2 T_{ab}.\label{feq}
\ee
Here $\kappa^2 = 8\pi G$ and $\Box$ is the covariant D'Alambertian. Any quantity with $,R$ in the subscript denotes 
derivative w.r.t $R$. \eqrf{feq} can also be written in the following form by using the expression for Einstein tensor $G_{ab}$
\be
f_{,R} G_{ab}-f_{,RR}\nabla_a\nabla_b R-f_{,RRR}\left(\nabla_a R\right)\left(\nabla_b R\right)+
            g_{ab}\left[\frac{1}{2}\left(R f_{,R}-f\right)+f_{,RR}\Box R+\right. \nonumber 
\ee
\be
\left.f_{,RRR}\left(\nabla R\right)^2\right] = \kappa^2 T_{ab},
\label{efeq}
\ee
where $\left(\nabla R\right)^2= g^{ab}\left(\nabla_a R\right)\left(\nabla_b R\right)$. The trace of the above field equation is given by
\begin{equation}
3f_{,RR}(R)\Box R-2f(R)+Rf_{,R}+3f_{,RRR}\left(\nabla R\right)^2=\kappa^2 T
\label{trace}
\end{equation}
which can be rewritten as
\be
\Box R = \frac{1}{3f_{,RR}}\left[\kappa^2 T-3f_{,RRR}\left(\nabla R\right)^2+2f - Rf_{,R}\right]
\label{Reqm}
\ee
The constant curvature solution of \eqrf{Reqm} in vacuum describes the de Sitter universe. It corresponds to the condition
\begin{equation}
2f(R_d)-R_df_{,R}(R_d)=0,
\label{desittercond}
\end{equation}
having de Sitter point denoted as $R_d$. A few de Sitter points for different values of $n$ and $b$ are listed 
in Table~\ref{tabl:desitter}. A de Sitter point $R_d$ is a stable point describing primordial and present vacuum energy dominated epoch if it satisfies the condition $F_{,R}(R_d)/F_{,RR}(R_d)>R_d$. All the de Sitter points given in Table~\ref{tabl:desitter} are stable de Sitter points.
The quantity $f_{,R}$ is identified with scalar field $\phi$ in $f(R)$ theory. We will discuss more about it in Sec.~\ref{sec:cursingularity}. The conditions for graviton to be of non-ghost nature and scalar field $\phi$ to be non techyonic are $f_{,R}>0$ and $f_{,RR}>0$ respectively. They are investigated for the model in \eqrf{model} and it is found that both the  conditions are satisfied by all parameter values mentioned in Table~\ref{tabl:desitter} for $R_d<R<\infty$. In the following sections, we will find that parameter values with $n\geq 1.75$ and $b\geq 1.021$ are constrained by local gravity tests and $n=1.75$ and $b=3.0$ is the best choice according to $Omh^2$ diagnostic.\\ Using \eqrf{Reqm} and \eqref{desittercond} we can define a potential 
\be
V(R) = -\frac{Rf(R)}{3}+\int^R f(R^\star) dR^\star, \label{modelpot} 
\ee
such that $\frac{dV(R)}{dR} = \frac{2f(R)-Rf_{,R}}{3} = 0$ for the  minimum at the de Sitter point. The potential
for some of the choices of parameters shown in Table~\ref{tabl:desitter} is shown in Fig.~\ref{fig:modelpot}.
\begin{table}[h!]
\centering
\begin{tabular}{|l|l|l|r|}
  \hline
  $n$           & $b$ & $\beta R_d$ \\
  \hline
   1.75     &                   1.021  & 2.4092    \\
   1.75     &                   1.1    & 2.7777    \\
   1.75     &                   1.5    & 4.2708    \\
   1.75     &                   2.0    & 5.9530    \\
   1.75     &                   3.0    & 9.1930    \\
   1.0      &                   2.0    & 5.1397    \\
   0.98     &                   5.0    & 14.6369   \\
  \hline
\end{tabular}
\caption{de Sitter points for different values of $b$ and $n$}
\label{tabl:desitter}
\end{table}

\subsection{Fixed Point Analysis} \label{FPA}
\begin{figure}[h!]
 \centering
 \includegraphics[width=0.5\textwidth]{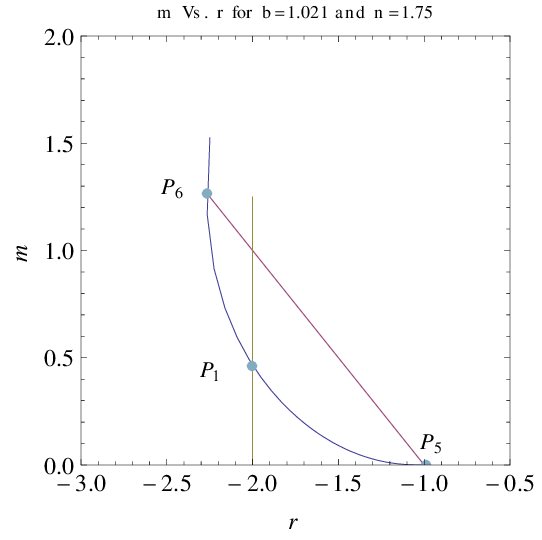}
 \caption{$m$ Vs. $r$. Red and yellow lines are $m=-r-1$ and $r=-2$ line respectively.}
 \label{fig:fixedpoint}
\end{figure}

\begin{figure}[h!]
 \centering
 \includegraphics[width=0.6\textwidth]{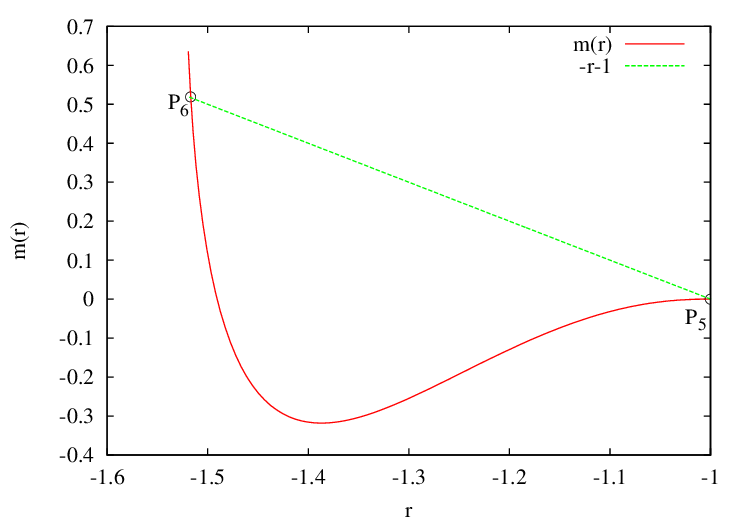}
 \caption{$m$ Vs. $r$ for Hu-Sawicki model for $n=4$. $c_1$ and $c_2$ are fixed using condition $|F_{,R_0}=0.01|$. Green line is $m=-r-1$ line.}
 \label{fig:fixedpoint1}
\end{figure}

It is necessary to carry out fixed point analysis in order to check whether given model can give rise to dark energy dominated era preceded by matter dominated era or not. In this section, we find the fixed points for the model \eqref{model} and also for Hu-Sawicki model proposed in \cite{Hu:2007nk}. Through this analysis, we show that our model belongs to different class than the Hu-Sawicki model does and therefore both models are phenomenologically distinguishable from each other. Here, we follow the procedure followed by many authors and originally proposed in \cite{Amendola:2006we}. Assuming background spacetime as a spatially flat FRW spacetime, field equations can be written as evolution equations in terms of $R$, $\dot R$, Hubble constant $H$ and matter density $\rho_m$. These evolution equations can be further written as a set of dynamical system equations by defining dimensionless variables $y_1\equiv-\dot{f_{,R}}/Hf_{,R},\;y_2\equiv-f/6f_{,R}H^2$, $y_3\equiv R/6H^2$ and $y_4\equiv \kappa^2\rho_r/3f_{,R}H^2$ together with the density parameters $\Omega_m\equiv 1-y_1-y_2-y_3-y_4$, $\Omega_r\equiv y_4$ and $\Omega_{DE}\equiv y_1+y_2+y_3$. Solving the set of dynamical system equations, in the absence of radiation ($y_4=0$), one can obtain several fixed points namely $P_1,P_2,P_3,$ $P_4,P_5$ and $P_6$. As described in \cite{Amendola:2006we}, among these fixed points, either $P_1$ or $P_6$ can give rise to late-time aceleration. The point $P_1$ is de Sitter point for which condition \eqref{desittercond} is satisfied. Basically, this dynamical system can be characterized by two quantities
\begin{equation}
m\equiv\frac{Rf_{,RR}}{f_{,R}},
\label{mexp1}
\end{equation}
\begin{equation}
r\equiv-\frac{Rf_{,R}}{f}=\frac{y_3}{y_2}.
\label{rexp1}
\end{equation}
The $m$ and $r$ for our model given in \eqref{model} can be written in terms of $x=\beta R$ as follows
\begin{equation}
m=\frac{2bn^2x^{3n-1}-bn(n-1)x^{n-1}(1+x^{2n})}{(1+x^{2n})(1+x^{2n}-bnx^{n-1})},
\label{mexp}
\end{equation}
\begin{equation}
r=\frac{-x-x^{2n+1}+bnx^n}{(1+x^{2n})(x-bArcTan(x^n))}.
\label{rexp}
\end{equation}
The function $f(R)$ for Hu-Sawicki model \cite{Hu:2007nk} and $m$ and $r$ are given by 
\begin{equation}
f(R)=R-\frac{c_1}{\beta}(\beta R)^n\left(1+c_2(\beta R)^n\right)^{-1}
\label{fhsr}
\end{equation}
\begin{equation}
m=\frac{(1+c_2x^n)^{-3}\left(-(2n-1)nc_1c_2x^{2n-1}+nc_1c_2x^{3n-1}\right)}{1-nc_1c_2x^{2n-1}(1+c_2x^n)^{-2}},
\label{mexp2}
\end{equation}
\begin{equation}
r=-\frac{-x-nc_1c_2x^{2n}(1+c_2x^n)^{-2}}{x-c_1x^n(1+c_2x^n)^{-1}}.
\label{rexp2}
\end{equation}
In Figures \ref{fig:fixedpoint} and \ref{fig:fixedpoint1}, $m(r)$ curve is plotted on $(r,m)$ plane for the function given in \eqrf{model} and \eqrf{fhsr} respectively. The points $P_5$ and $P_6$ exist on $m=-r-1$ line. The saddle matter era is realized by fixed point $P_5$. 
The condition for the existence of saddle point $P_5$ is $m(r)=+0,\;dm/dr>-1,$ at $r=-1$. If we take limit $\beta R\rightarrow\infty$ in expressions of $r$ in \eqrf{rexp} and \eqrf{rexp2}, we can obtain $r\rightarrow -1$. Calculating $dm/dr$ from $(dm/dR)/(dr/dR)$ and taking $R\rightarrow\infty$ limit in it, one can end up with $dm/dr(r=-1)\geq -1$. This says that $P_5$ is a saddle point showing existence of saddle matter era in our model and also in Hu-Sawicki model at very large curvature. The stability condition for the point $P_6$ is $m=-r-1,(\sqrt3-1)/2<m<1$ which is satisfied in Hu-Sawicki model but not in our model as evident from the Figures \ref{fig:fixedpoint} and \ref{fig:fixedpoint1}. Lastly, the de Sitter point $P_1$ for the model \eqref{model}, as we have already obtained in Sec.~\ref{subsec:desitter}, is a stable point. This can be also proved from its stability condition $(r=-2,\;0<m\leq 1)$ which is clearly satisfied by our model. Therefore, we can conclude that the model given in \eqrf{model} gives the evolution of our universe from saddle matter era (point $P_5$) to a de Sitter state (point $P_1$) through blue curve shown in Fig.~\ref{fig:fixedpoint} while $m=0$ line shows the $\Lambda$CDM evolution. The universe given by Hu-Sawicki model evolves from saddle matter era (point $P_5$) to a late-time accelerating phase (point $P_6$) through red curve shown in Fig.~\ref{fig:fixedpoint1}. In the language of \cite{Amendola:2006we}, any satisfacory $f(R)$ model belongs to either class II:  the matter epoch is followed by a de Sitter acceleration (connecting $P_5$ to $P_1$)  or class IV: the matter epoch is followed by a nonphantom accelerated attractor (connecting $P_5$ to $P_6$). From this analysis, we can say that our model given in \eqref{model} belongs to class II and Hu-Sawicki model given in \eqref{fhsr} belongs to class IV.

\section{Fifth Force Constraint }\label{sec:fifthforce}
It is well-known that the $f(R)$ theory has an extra scalar degree of freedom in addition to usual graviton. Since the scalar field is a dynamical degree of freedom, it gives rise to a propagating fifth-force which is the cause of the late-time acceleration. But, it must be suppressed at the local gravity scales in order to evade experimental results of solar system tests and EP(Equivalence Principle) violation tests. The condition for the suppression of fifth force on local gravity scales is given by $m_{\phi}^2L_s^2>>1$, where $m_{\phi}$ is the mass of the scalar field and $L_s$ is the typical length scale involved in experiment \cite{Olmo:2005hc,Olmo:2005zr}. One can calculate the value of $m_{\phi}^2$ for our model given in \eqref{model}, which comes out to be of the order of $10^{-51} m^{-2}$. Considering $L_s=1m$, the above condition can be seen to be completely violated.\\
In the Einstein frame, the scalar field of an $f(R)$ theory is a chameleon-like field which exhibits chameleon mechanism given by \cite{Khoury:2003aq}. The name owes to the fact that the "chameleon" field blends with the environment in the highly dense background and becomes invisible to any experimental searches for fifth-force. This is because the mass of the field depends on the background density. It can be shown as following. The action in \eqrf{action} can be written, via a conformal transformation, in Einstein frame as
\begin{equation}
\small{
S = \int {d^4x \sqrt{-\tilde g}\left[\frac{\tilde R}{2\kappa^2} - \frac{(\tilde\nabla\psi)^2}{2} - V_E(\psi) + \mathfrak{L}_m(\tilde g_{\mu\nu}e^{-\frac{2}{\sqrt{6}}\kappa\psi})\right]}.
}
\label{11}
\end{equation}
Here, all quantities having tilde are defined in Einstein frame. The Einstein frame scalar field $\psi$ is related to function $f(R)$ by the relation $\kappa\psi=\sqrt{3/2}\ln{f_{,R}}$. Let us consider the spherically symmetric body with radius $\tilde r_c$. Varying action in \eqref{11} w.r.t. $\psi$ we obtain the field equation for $\psi$ as
\begin{equation}
\frac{d^2\psi}{d\tilde r^2}+\frac{2}{\tilde r}\frac{d\psi}{d\tilde r}-\frac{dV_{eff}}{d\psi}=0.
\label{fepsi}
\end{equation}
The effective potential $V_{eff}$ is given by
\begin{equation}
V_{eff}(\psi)=V_E(\psi)+e^{-\frac{2}{\sqrt 6}\kappa\psi}\rho^*.
\label{effpotein}
\end{equation}
Because of the interplay of two terms self-interaction potential $V_E$ and conformal coupling to the matter, the effective potential possesses the global minimum. By solving the dynamical equation \eqref{fepsi}, one can find that the scalar field $\psi$ is frozen at the potential minimum with a very small magnitude and its mass is very high. Thus, its contribution to the field outside is negligible except for a thin shell regime near the surface of the object. The thin-shell condition is given by
\begin{equation}
\frac{\delta\tilde{r}_c}{\tilde{r}_c}=-\frac{\psi_{out}-\psi_{in}}{\sqrt 6\Phi_c}.
\label{thinshellcond}
\end{equation}
Where, $\Phi_c$ is the gravitational potential on the surface of the test body (Sun/Earth) and $\delta\tilde{r}_c$ is the thickness of the thin shell. Here, $\psi_{out}$ and $\psi_{in}$ are value of the field at the minima of $V_{eff}$ outside and inside the body respectively. If $\psi_{in}<<\psi_{out}$ the the thin-shell condition written in Eq.\eqref{thinshellcond} can be reduced to
\begin{equation}
\left|\psi_{out}\right|\simeq\sqrt{6}\Phi_c\frac{\delta\tilde{r}_c}{\tilde{r}_c}.
\label{gencondpsiout}
\end{equation}
\begin{figure}[h!]
 \centering
 \includegraphics[width=0.5\textwidth]{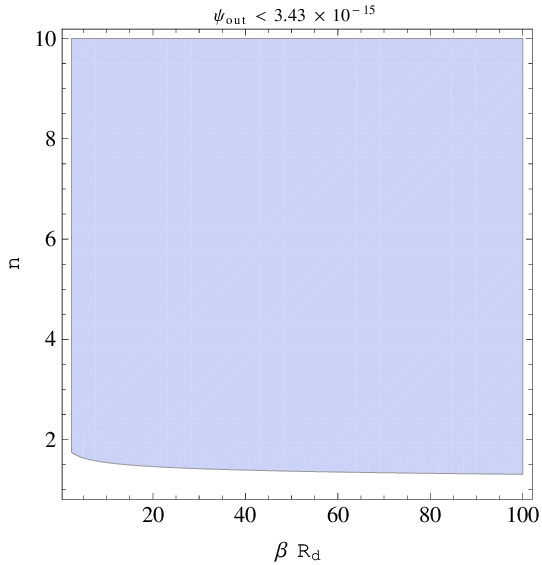}
 \caption{Shaded region is the allowed region for different values of parameter $n$ and $\beta R_d$.}
 \label{fig:fifthforcenewarctan}
\end{figure}
The experimental results of the solar system tests and the EP violation \cite{Will:2014xja,Capozziello:2007eu} tests put the constraints on the r.h.s. of the Eq. \eqref{gencondpsiout}:
\begin{eqnarray}
\label{eq:phiconstraint}
&\lesssim& \left \{ \begin{array}{rl}
 5.97\times 10^{-11} &\textrm{(Solar system test)}, \\
 3.43\times 10^{-15} &\textrm{(EP test)}.
\end{array} \right .
\label{ffcond}
\end{eqnarray}
For the model given in \eqref{model}, $\psi_{out}$ can be written as 
\begin{equation}
\small{
\psi_{out}\approx \frac{\sqrt 6}{2}\left.F_{,R}\right|_{R=\rho_{out}}=\frac{\sqrt 6}{2}\left[\frac{(-nb)(\beta\rho_{out})^{n-1}}{\{1+(\beta\rho_{out})^{2n}\}}\right].
}
\label{condpsiout}
\end{equation}
Here, we have considered the approximation that $R>>1/\beta$. Let us put $\beta\rho_{out}=\beta R_d\frac{\rho_{out}}{R_d}$. Since $\rho_{out}\simeq10^{-24}g/cm^3$ and $R_d\simeq\rho_c\sim10^{-29}g/cm^3$ one can write $\beta\rho_{out}\simeq\beta R_d\times10^5$ in \eqrf{condpsiout}. 
\begin{equation} 
\psi_{out}\approx \frac{\sqrt 6}{2}\left[-nb\frac{(\beta R_d\times10^5)^{n-1}}{\{1+(\beta R_d\times10^5)^{2n}\}}\right]
\label{condpsiout2}
\end{equation}
From the de Sitter condition in \eqrf{desittercond}, parameter $b$ can be written in terms of $n$ and $\beta R_d$ as
\begin{equation}
b=\frac{-\beta R_d}{-2Arctan(\beta R_d)^n+\frac{n(\beta R_d)^n}{1+(\beta R_d)^{2n}}}.
\label{expb}
\end{equation}
Plugging \eqrf{expb} into \eqrf{condpsiout2}, one ends up with the expression
\be 
\left|\psi_{out}\right|\approx \left|\frac{-n\sqrt 6}{2}\left(\frac{-\beta R_d}{-2Arctan(\beta R_d)^n+\frac{n(\beta R_d)^n}{1+(\beta R_d)^{2n}}}\right)\times\frac{(\beta R_d\times10^5)^{n-1}}{\{1+(\beta R_d\times10^5)^{2n}\}}\right|
\label{condpsiout3}
\ee
With the help of \eqrf{condpsiout3} and \eqrf{eq:phiconstraint} we can put the constraint on our model parameters to evade the fifth-force constraint. Fig.~\ref{fig:fifthforcenewarctan} exhibits allowed regions for different values of parameter $n$ and $\beta R_d$.
From the Fig.~\ref{fig:fifthforcenewarctan}, one can conclude that $b \geq 1.021$ and $n\geq  1.75 $ to evade the local gravity test. The corresponding de Sitter point $\beta R_d$ has to be $\geq 2.4 $.


\section{Curvature Singularity}\label{sec:cursingularity}
The trace equation \eqref{trace} can also be written as
\begin{equation}
3\Box F_{,R}(R)-2F-R+RF_{,R}(R)=\kappa^2 T
\label{trace1}
\end{equation}
Rewriting the term $\Box F_{,R}=\Box\phi$, where $\phi$ is the scalar field in the Jordan frame, one can write equation of motion for $\phi$ as 
\begin{equation}
\Box\phi=\frac{dV_J}{d\phi}+\frac{\kappa^2}{3}T,
\label{oscillations}
\end{equation}
where
\begin{equation}
\frac{dV_J}{d\phi}=\frac{1}{3}(R+2F-RF_{,R}).
\label{dvjbydphi}
\end{equation}
Here, $V_J$ is the scalar field potential in the Jordan frame. For the model written in \eqref{model}, $dV_J/d\phi$ can be written as
\be
\small{
\frac{dV_J}{d\phi}=\frac{1}{3}\left[-2\frac{b}{\beta}\tan^{-1}(\beta R(\phi))^n+R+\frac{b}{\beta}\frac{n(\beta R(\phi))^n}{1+(\beta R(\phi))^{2n}}\right]
}
\label{dvjbydphi1}
\ee
The dynamics of the scalar field $\phi$ depends solely on the potential $V_J$ in vacuum. While in the presence of matter, the term $\kappa^2 T$ will also play a role. In this case, the equation of motion can be written as
\begin{equation}
\Box\phi=\frac{\partial V_J^{eff}}{\partial\phi},
\label{oscillations2}
\end{equation}
where
\begin{equation}
\frac{\partial V_J^{eff}}{\partial\phi}=\frac{1}{3}(R+2F-RF_{,R}+\kappa^2 T).
\label{dvjeffbydphi}
\end{equation}

\begin{figure*}
  \centering
   $
   \begin{array}{c c}
   \includegraphics[width=0.48\textwidth]{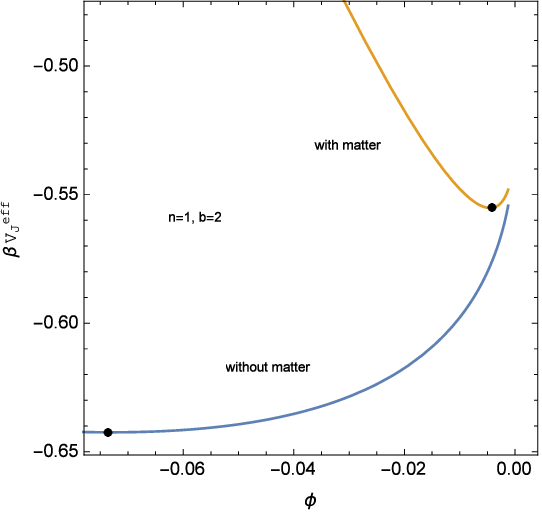} & 
   \includegraphics[width=0.48\textwidth]{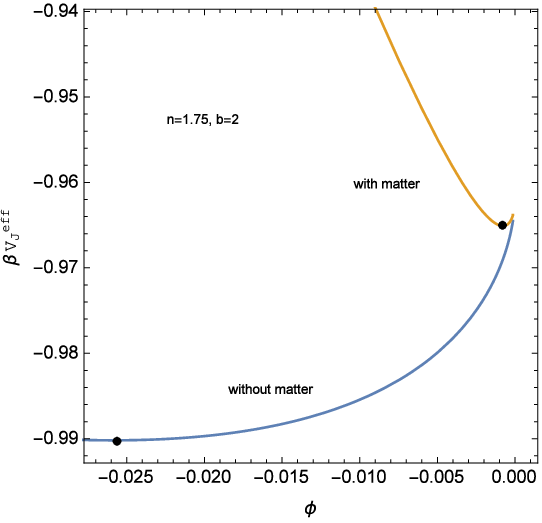}
   \end{array}
   $ 
  \caption{$\beta V_J^{eff}$ Vs. $\phi$. Left figure is for $n=1,\;b=2$ and right figure is for $n=1.75,\;b=2$.}
  \label{fig:jordanpot}
\end{figure*}
If we integrate the \eqrf{dvjbydphi1} with respect to $\phi$, we obtain $V_J(R(\phi))$. The minimum of the potential $V_J$ corresponds to the de Sitter point in the vacuum about which the field $\phi$ oscillates. This minimum can be at a finite distance in $\phi$ and with a finite potential difference from the singular point $\phi=0$. At the singular point $\phi=0$, the scalar curvature $R$ becomes infinity and therefore it is called curvature singularity. \\
The plots for $V_J^{eff}(\phi)$ with matter and without matter are shown for different values of parameters $n$ and $b$ in Fig.~\ref{fig:jordanpot}. The black dots are minima of the potential $V_J^{eff}$. It can be easily seen that the minimum moves closer to the singularity in the presence of matter thus making the scalar field more vulnerable to meet the singularity. The minimum is at a closer distance for larger values of $n$ as evident from the Fig.~\ref{fig:jordanpot}. Therefore the Arctan model is safer than our model in \eqref{model} but not free from fatal curvature singularity at all. But, the curvature singularity can be cured by adding $R^2$ term to the Lagrangian since it can increase the potential $V_J$ to the infinity as $\phi$ approaches the singularity for fine-tuned parameter values as suggested in \cite{Appleby:2009uf}.


\section{Cosmological Evolution}\label{sec:cosmoevol}
In this section, we discuss cosmological implications of the model given in \eqrf{model}. We consider homogeneous and isotropic FRW universe described by the metric
\be
\small{
ds^2 = -dt^2 + a^2(t)\left[\frac{dr^2}{1-K r^2}+r^2\left(d\theta^2+\sin^2\theta d\phi^2\right)\right]
}
\ee 
Using \eqrf{Reqm} with above spacetime we get
\be
\small{
     \ddot R = -3H \dot R -\frac{1}{3f_{,RR}}\left[3f_{,RRR}\dot R^2+2f-f_{,R}R+\kappa^2 T\right] 
}
\label{tracefrw}
\ee 
The cosmological evolution can be obtained by solving field equation \eqref{efeq} with FRW metric and diagonal energy-momentum tensor. The equations governing the scale factor and Hubble constant $H$ for flat FRW universe are given as
\be
H^2  + \frac{1}{f_{,R}}\left[f_{,RR}H\dot R-\frac{1}{6}\left(f_{,R}R-f\right)\right]= 
-\frac{\kappa^2 T^t_t}{3f_{,R}}, \label{freq1}
\ee
\be
  \dot H  =  -H^2+\frac{1}{f_{,R}}\left(f_{,RR}H \dot R +\frac{f}{6}+\frac{\kappa^2T^t_t}{3}\right) \label{freq2},
\ee
where $H = \frac{\dot a}{a}$. The expression for the Ricci scalar in terms of Hubble constant directly obtained from the metric is 
given by
\be
R = 6 \left(\dot H + 2 H^2 + \frac{K}{a^2}\right)\label{rs}
\ee

The energy-momentum tensor considered for cosmological evolution has three contributions i.e. baryon, radiation, and dark matter.
The conservation equation $\nabla_a T^{ab} = 0$, satisfied by each component separately, leads to the following equation,
\be
\dot \rho_T +3 H \left(\rho_T+p_T\right) = 0.\label{eqofcon}
\ee
Here the total energy density is $\rho_T = \rho_{bar}+\rho_{DM}+\rho_{rad}$. The above equation can be integrated by using
$p_{bar},\, p_{DM} = 0$ and $p_{rad} = \rho_{rad}/3$ and the energy density can be expressed in terms of scale factor as
\be
\rho_T = \frac{\rho_{bar}^0+\rho_{DM}^0}{\left(a/a_0\right)^3}+\frac{\rho^0_{rad}}{\left(a/a_0\right)^4},\label{ed}
\ee
where the knotted quantities indicate their values today. The time-time component of the energy-momentum tensor and its trace 
appearing in equations \eqref{tracefrw}, \eqref{freq1} and  \eqref{freq2} can be written in terms of energy density and pressure as $T^t_t = -\rho_T$  and 
$T = -\left(\rho_{bar}+\rho_{DM}\right)$.

Now to obtain the equation of state of $f(R)$ gravity one can define the energy density $\rho_X$ such that the Friedmann equation
(\ref{freq1}) looks like
\be
H^2 = \frac{\kappa^2}{3}\left(\rho+\rho_X\right) \label{freq1n}.
\ee
\begin{figure*}
  \centering
   $
   \begin{array}{c c}
   \includegraphics[width=0.48\textwidth]{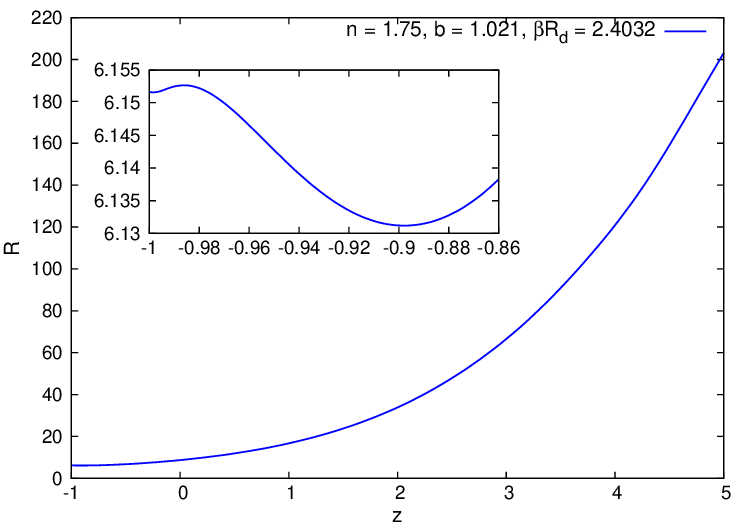} & 
   \includegraphics[width=0.48\textwidth]{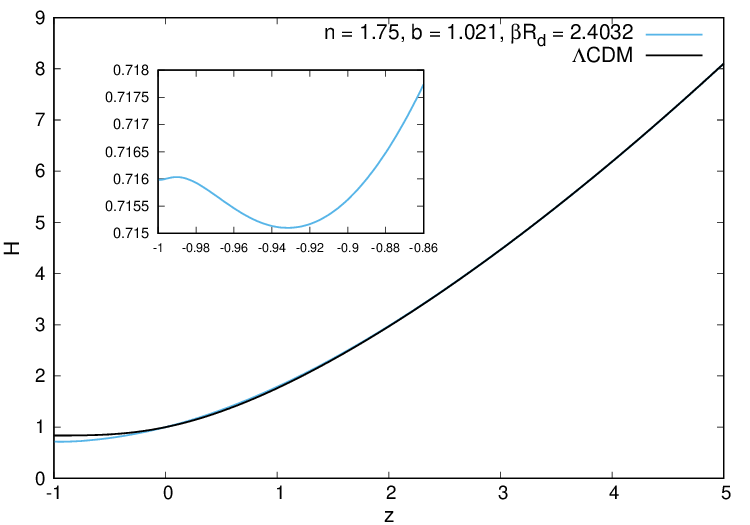}
   \end{array}
   $ 
  \caption{Left figure is $R$ Vs. $z$ and right figure is $H$ Vs. $Z$}
  \label{fig:model1RH}
\end{figure*}

\begin{figure*}
  \centering
   $
   \begin{array}{c c}
   \includegraphics[width=0.48\textwidth]{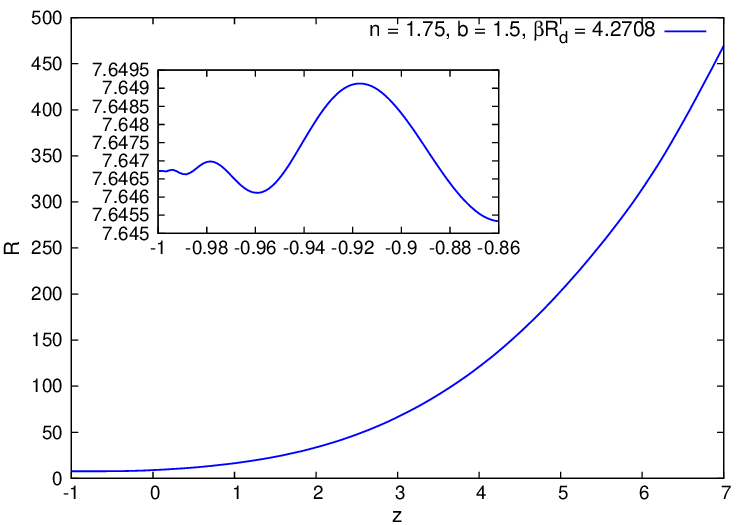} & 
   \includegraphics[width=0.48\textwidth]{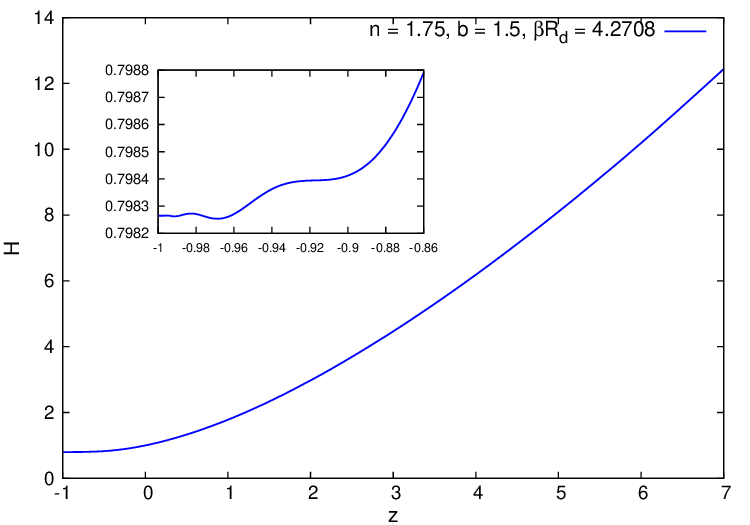}
   \end{array}
   $ 
  \caption{Left figure is $R$ Vs. $z$ and right figure is $H$ Vs. $Z$}
  \label{fig:model3RH}
\end{figure*}

\begin{figure*}
  \centering
   $
   \begin{array}{c c}
   \includegraphics[width=0.48\textwidth]{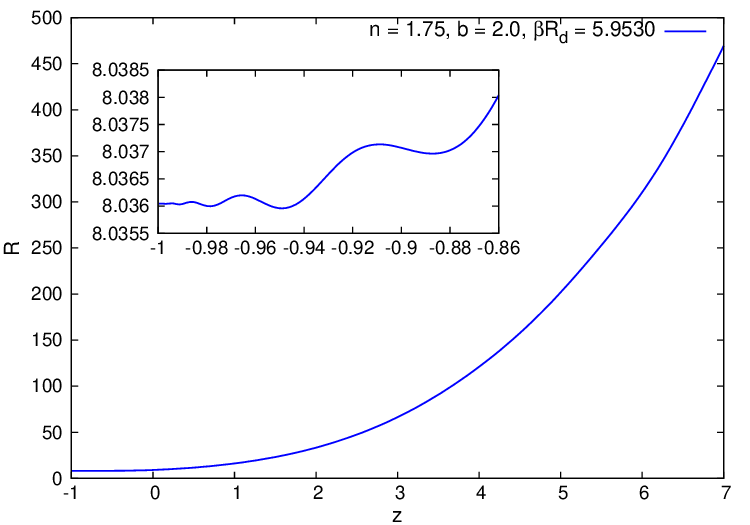} & 
   \includegraphics[width=0.48\textwidth]{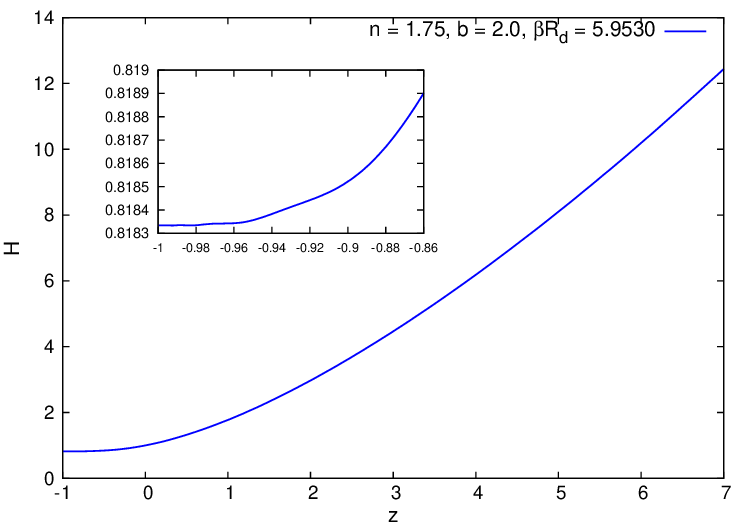}
   \end{array}
   $ 
  \caption{Left figure is $R$ Vs. $z$ and right figure is $H$ Vs. $Z$}
  \label{fig:model4RH}
\end{figure*}

\begin{figure*}
  \centering
   $
   \begin{array}{c c}
   \includegraphics[width=0.48\textwidth]{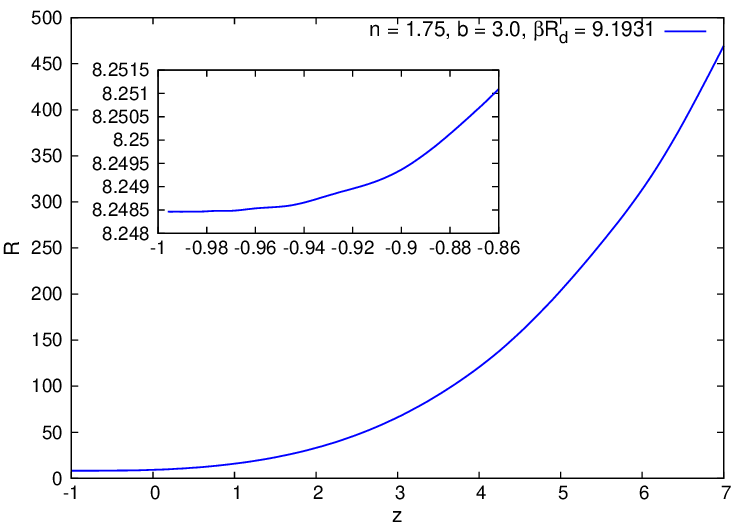} & 
   \includegraphics[width=0.48\textwidth]{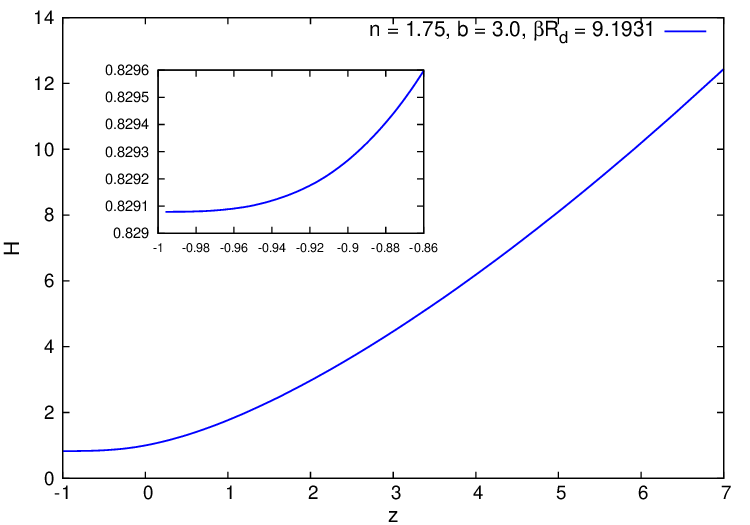}
   \end{array}
   $ 
  \caption{Left figure is $R$ Vs. $z$ and right figure is $H$ Vs. $Z$}
  \label{fig:model5RH}
\end{figure*}

\begin{figure}
 \centering
 \includegraphics[width=0.5\textwidth]{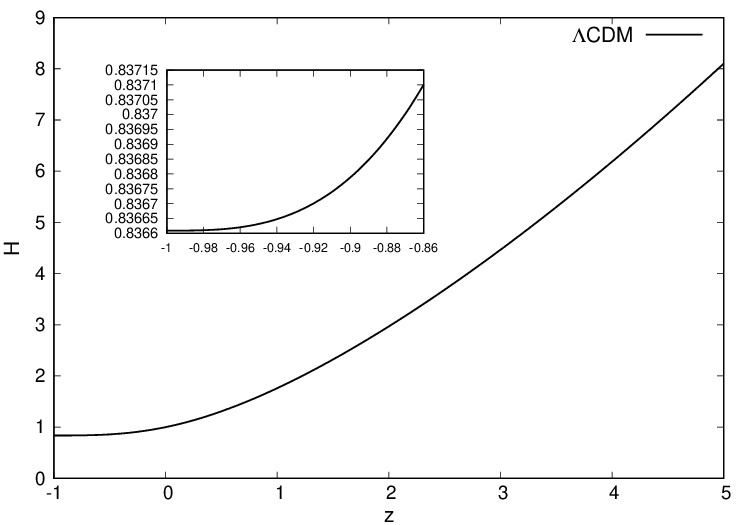}
 \caption{$H$ Vs. $z$ for $\Lambda$CDM.}
 \label{fig:lcdmH}
\end{figure}

Similarly the pressure $p_X$ can also be defined in a way so that the \eqrf{freq2} will become
\be
\dot H + H^2 = -\frac{\kappa^2}{6}\left(\rho+\rho_X+3(p_{rad}+p_X)\right) \label{freq2n}.
\ee
Now using the above two equations along with \eqrf{freq1} and \eqrf{freq2} the energy density and pressure for $f(R)$ fluid can be
written as
\be
\small{
\rho_X = \frac{1}{\kappa^2f_{,R}}\left[\frac{1}{2}\left(f_{,R}R - f\right)-3 f_{,RR}H \dot R+\kappa^2\rho\left(1-f_{,R}\right)
\right],
}
\label{rhox}
\ee
\be
\small{
p_X =-\frac{1}{3\kappa^2f_{,R}}\left[\frac{\left(f_{,R}R + f\right)}{2}+3 f_{,RR}H \dot R - \kappa^2\left(\rho-3 p_{rad}f_{,R}\right)\right].
}
\label{pressurex}
\ee
With these expressions of energy density and pressure, we can write the equation of state for $f(R)$ as
\be
w_X = \frac{p_X}{\rho_X}.\label{eqs1}
\ee
The equation of state $w_X$ can also be written in the following form by using Eqns.~(\ref{freq1n}), (\ref{freq2n}), and (\ref{rs}) as
\be
w_X = \frac{3 H^2-3 \kappa^2 p_{rad}-R}{3\left(3H^2-\kappa^2\rho\right)}.\label{omegax1} 
\ee
To obtain the behavior of $R$ and $H$ w.r.t redshift $z$, we solve Eqns.\eqref{tracefrw}, \eqref{freq1}, \eqref{freq2}, and \eqref{rs} numerically using the method described in \cite{Jaime:2012gc}. To integrate these differential equations we choose
\be
\alpha= \ln (a/a_0)
\ee
 as an independent variable instead of $t$ since $\alpha\rightarrow -\infty$ as $a\rightarrow 0$ is well kept far from the regime of integration. Also, it is easy to get the quantities $R$ and $H$ in terms of redshift $z=e^{-\alpha}-1$. 
Eqns.~\eqref{tracefrw}, \eqref{freq1}, \eqref{freq2} and \eqref{rs} can be expressed
in terms of $\alpha$ as
\be
R^{\prime\prime}=-R^\prime\left(1+\frac{R}{6 H^2}\right)-\frac{1}{3f_{,RR}H^2}\left[3f_{,RRR}H^2R^{\prime2}+2f- f_{,R}R+\kappa^2T\right],
\label{deq1}
\ee
\be
H^\prime  =   -2 H + \frac{R}{6 H},
\label{deq2}
\ee
\be
H^2 + \frac{1}{f_{,R}}\left[f_{,RR}H^2 R^\prime  - \frac{1}{6}\left(f_{,R}R-f\right)\right] = - \frac{\kappa^2T^t_t}{3f_R},
\label{deq3}
\ee
\be
H^\prime = - H + \frac{1}{f_{,R}H}\left(f_{,RR}H^2 R^\prime+\frac{f}{6}+\frac{\kappa^2 T^t_t}{3}\right),
\label{deq4}
\ee
where ' denotes the derivative w.r.t. $\alpha$.

To numerically integrate the above differential equations we can use either \eqrf{deq2} or \eqref{deq4} for $H$. To check the 
consistency of the code we have used both. \eqrf{deq3} is modified Hamiltonian constraint and can be used to determine the 
accuracy of the code. The initial conditions for $R$ and $H$ are taken as described in \cite{Jaime:2012gc}.

The behavior of Ricci scalar $R$ and Hubble constant $H$ is depicted in Figs.~\ref{fig:model1RH} 
to \ref{fig:model5RH} for the choice of the parameters given in Table \ref{tabl:desitter}. 
As ahown in the inset figures which is a zoomed, The Ricci scalar oscillates around its value at the de Sitter point today, 
which can be seen by applying linear perturbations (see Fig.~\ref{fig:modelpot}) around the de Sitter minimum.
The Hubble constant today also oscillates. From Fig.~\ref{fig:model5RH} we see that the
oscillations are not significant for larger value of $b$ with $n = 1.75$.  In Fig.~\ref{fig:model1RH}, evolution of $H(z)$ in $\Lambda$CDM is also shown for the comparision. 

\begin{figure}[h!]
 \centering
 \includegraphics[width=4.8cm,height =7.6cm,angle = -90]{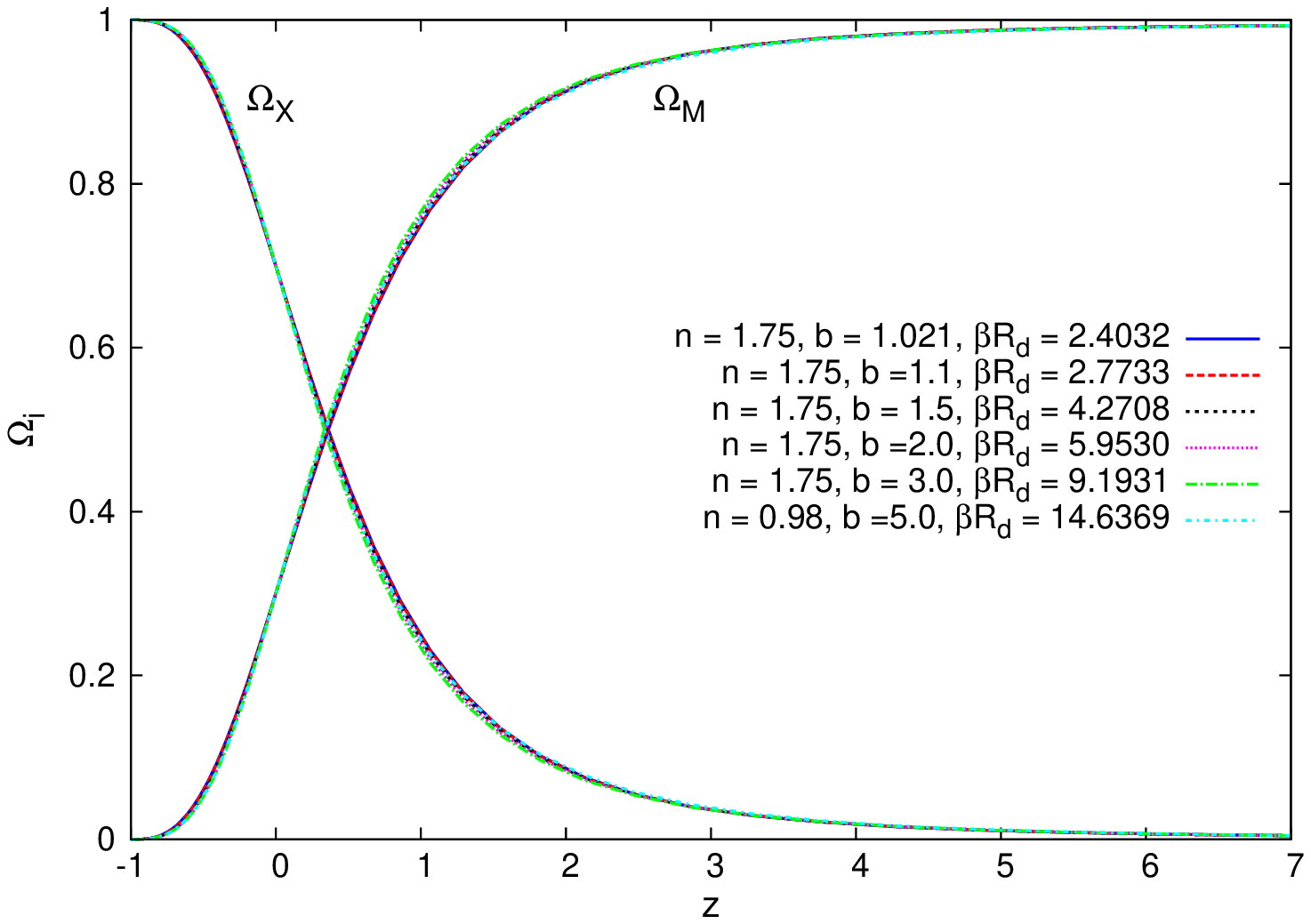}
 \caption{$\Omega_i$ vs $z$}
 \label{fig:omegavsz}
\end{figure}

\begin{figure}[h!]
 \centering
 \includegraphics[width=4.8cm,height =7.6cm,angle = -90]{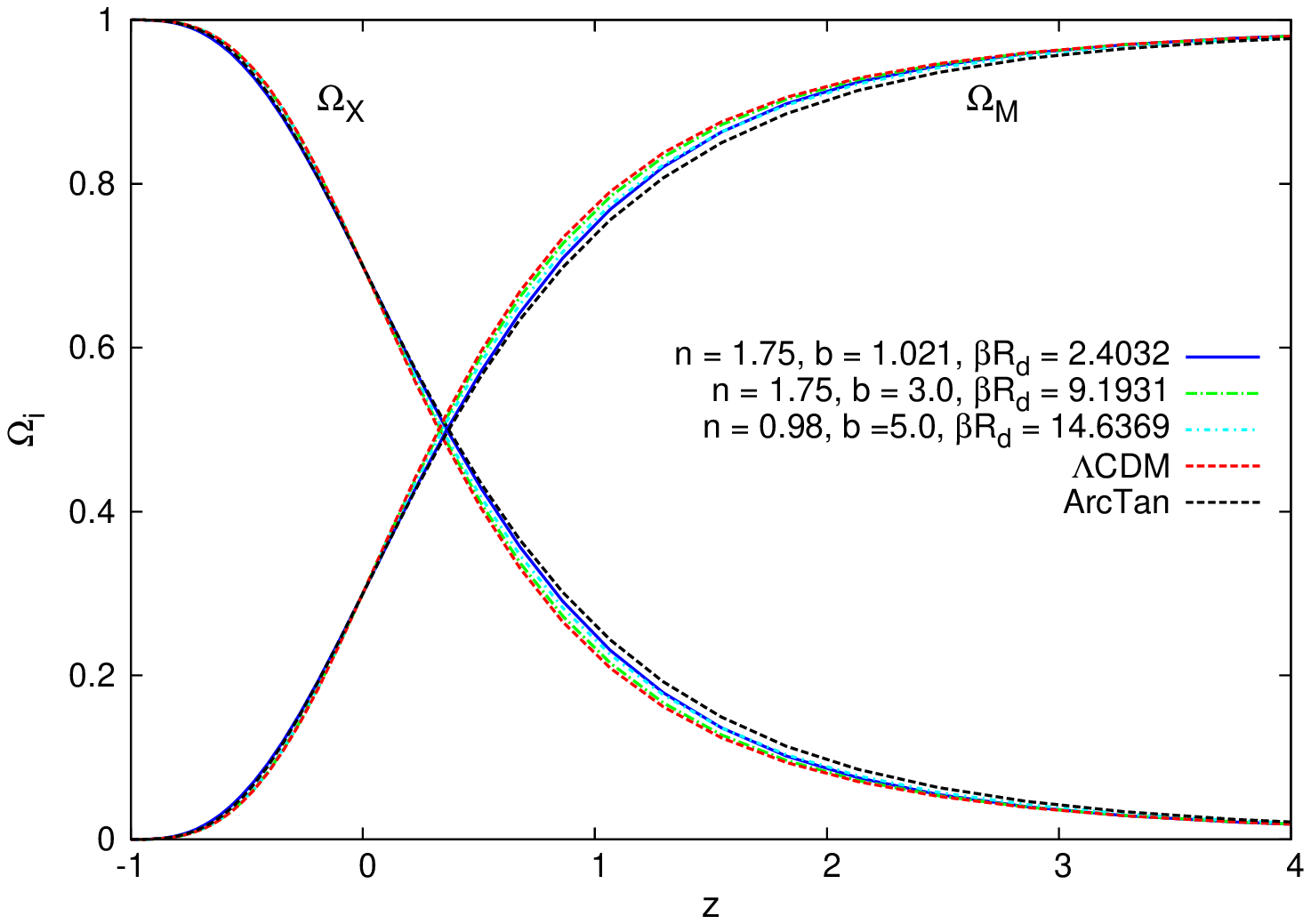}
 \caption{$\Omega_i$ vs $z$}
 \label{fig:omegavsz2}
\end{figure}

The dimensionless densities for different species can be expressed as
\be
\Omega: = \Omega_{rad} + \Omega_{bar}+\Omega_{DM} +\Omega_X = 1,
\ee
where $\Omega_i = \frac{\kappa \rho_i}{3 H^2}$. The density for the $f(R)$ model is given by \eqref{rhox}.
The variation of matter density $\Omega_M$ and the density of $f(R)$ model (\ref{model}) 
$\Omega_X$ w.r.t $z$ is shown in Figs.~\ref{fig:omegavsz} and \ref{fig:omegavsz2} for the values  of $n$ and $b$ given in 
Table \ref{tabl:desitter}. 
We have also plotted the same densities for the $\Lambda$CDM model for reference. The various
density parameters $\Omega_i$ for the $\Lambda$CDM model can be expressed in terms of their values today 
as
\be
\small{
\Omega_i^{\Lambda CDM} = \frac{\Omega_i^{0\, \Lambda CDM} \bar{a}^I}
{\left[\left(\Omega^{0\, \Lambda CDM}_{bar}+\Omega^{0\, \Lambda CDM}_{DM}\right)\bar{a}^{-3}
+\Omega^{0\, \Lambda CDM}_{rad}\bar{a}^{-4}+\Omega_\Lambda^0\right]}.
}
\label{omegalcdm}
\ee 
 Here $i$ represents baryon, dark matter, radiation and $\Lambda$ and $I = -3,\, -3,\, -4,\, 0$ for 
these species respectively. It is clear from the Figs.~\ref{fig:omegavsz} and \ref{fig:omegavsz2} that the variation of 
$\Omega_i$ w.r.t $z$ is not much different for all the choices of the parameters $n$ and $b$. It is also clear from the
Figs.~\ref{fig:omegavsz} and \ref{fig:omegavsz2} that the $f(R)$ model (\ref{model}) 
presented in this work exhibit sufficiently long matter dominated era as predicted by fixed point analysis presented in Sec. \ref{FPA}. For larger redshift $\Omega_X$ behaves in a similar way as $\Omega_{\Lambda}$ so we expect a 
radiation era in the early universe similar to $\Lambda$CDM model so that the predictions of 
Nucleosynthesis are not spoiled.

\begin{figure}[h!]
 \centering
 \includegraphics[width=7.6cm,height =4.8cm,angle = 0]{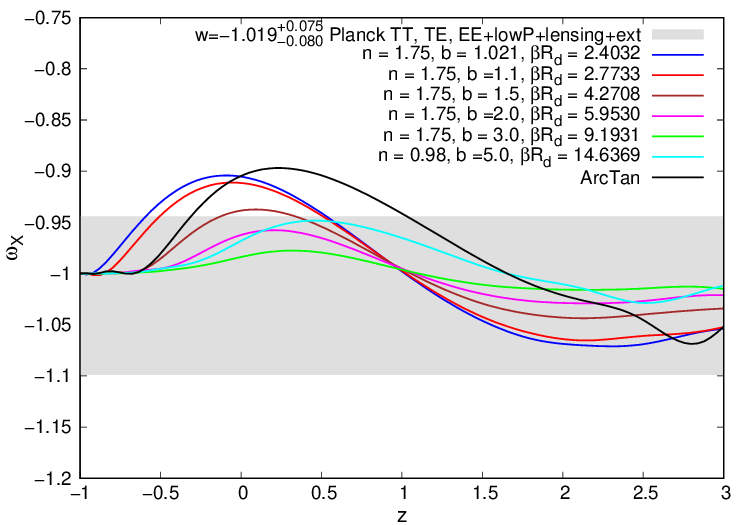}
 \caption{$\omega_X$ vs $z$. The shaded region represents $95\%$ constraints with Planck TT, TE, EE+lowP+lensing+ext}
 \label{fig:wxvsz}
\end{figure}
\begin{figure}[h!]
 \centering
 \includegraphics[width=4.8cm,height =7.6cm,angle = -90]{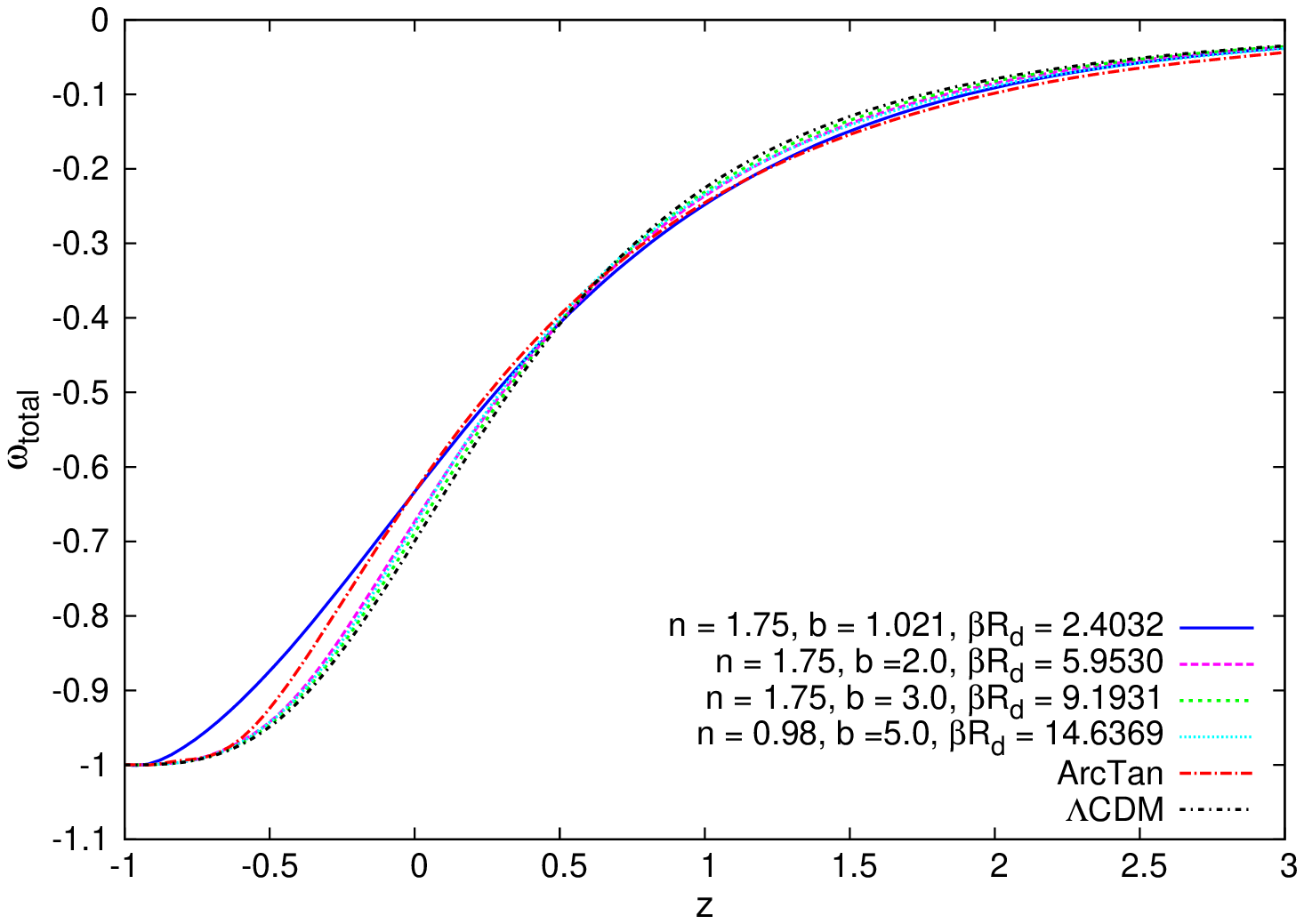}
 \caption{$\omega_{total}$ vs $z$}
 \label{fig:wtotalvsz}
\end{figure}
Figs.~\ref{fig:wxvsz} and \ref{fig:wtotalvsz} show the behavior of equation of state of the $f(R)$ component of the fluid
given by (\ref{model}) and the  equation of state of the total fluid. In Fig.~\ref{fig:wxvsz}, one can note that the values of equation of state for parameters with values $n=1.75$ and $b=2,3$ lie in the shaded region which shows the $95\%$ constraints with Planck TT, TE, EE+lowP+lensing+ext.  
From Fig.~[\ref{fig:wxvsz}] we see that the equation of state for the $f(R)$ model (\ref{model}) exhibits oscillation around phantom divide $\omega = -1$ for all choice of parameters. The existence of matter era for higher redshifts can also be seen from the
behavior of $\omega_{total}$ as depicted in Fig.~\ref{fig:wtotalvsz}.

\section{Luminosity distance and distance modulus}\label{sec:lddm}
\begin{figure}[h!]
 \centering
 \includegraphics[width=4.8cm,height =7.6cm,angle = -90]{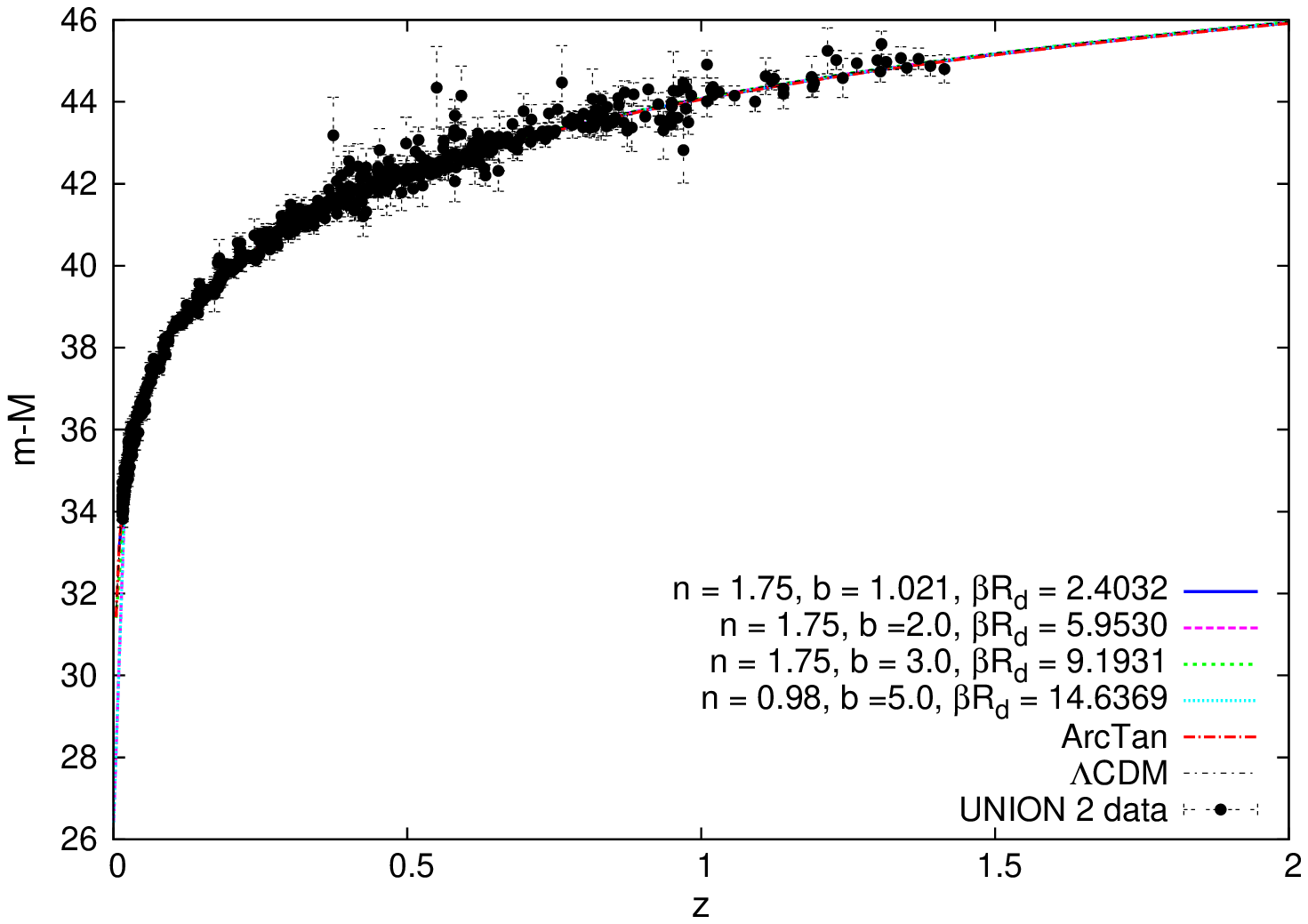}
 \caption{Distance modulus for the model (\ref{model}) along with the UNION 2 data of 
\cite{0004-637X-716-1-712} }
 \label{fig:spnvdata1}
\end{figure}

\begin{figure}[h!]
 \centering
 \includegraphics[width=4.8cm,height =7.6cm,angle = -90]{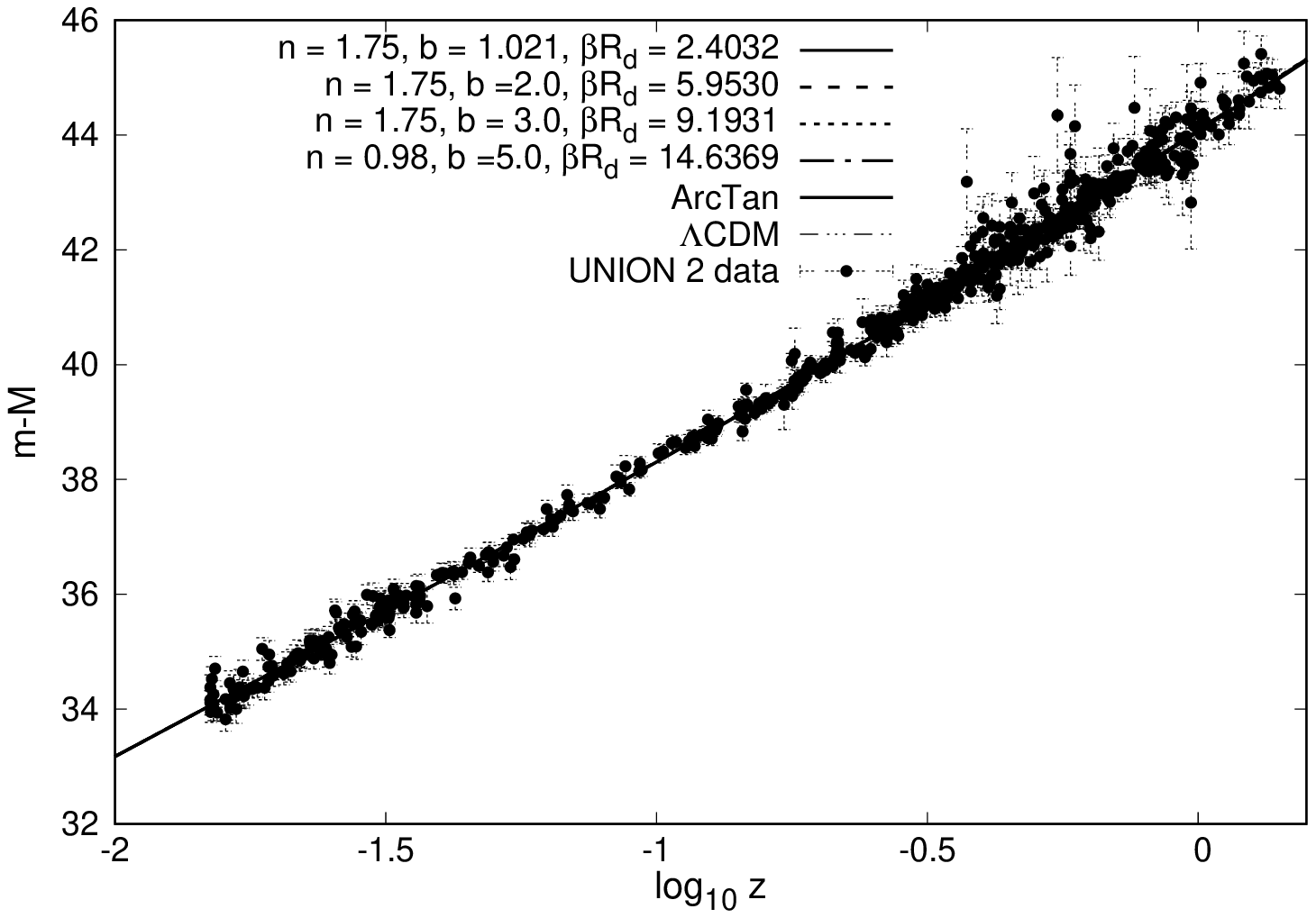}
 \caption{Distance modulus for the model (\ref{model}) along with the UNION 2 data of 
\cite{0004-637X-716-1-712} }
 \label{fig:spnvdata2}
\end{figure}

The observational evidence for dark energy comes from measuring luminosity distance to supernovae (SNIa). 
We can use the same supernovae data to constrain our new model of $f(R)$ theory. The luminosity distance is given by
\be
d_L = \frac{\zeta(\bar{a})}{\bar{a}}, \label{lmds}
\ee
where $\bar{a}= \frac{a}{a_0}$ and 
\be
\zeta = c H_0^{-1}\int_{\bar{a}}^1\frac{d\bar{a}^\star}{\bar{a}^{\star 2}\bar{H}({\bar{a}^\star})}. \label{zeta}
\ee
Here speed of light has been introduced explicitly to compute distance in units of Mpc and $\bar{H}= \frac{H}{H_0}$.
In order to compute $\zeta$ by the numerical method described in \cite{Jaime:2012gc}, we can transform the expression (\ref{zeta}) in terms
of differential equation 
\be
\frac{d\bar{\zeta}}{d\bar{a}} = -\frac{1}{\bar{a}^2\bar{H}(\bar{a})}, \label{dezeta}
\ee 
where $\bar{\zeta}= \frac{\zeta}{\left(c H_0^{-1}\right)}$ is  dimensionless and the above differential equation in terms of
variable $\alpha =\ln (\bar{a})$ can be written as
\be
\bar{\zeta}^\prime = -\frac{e^{-\alpha}}{\bar{H}}. \label{dezeta1}
\ee
The solution for the above differential equation can be obtained simultaneously with the field equations. The initial 
condition for $\zeta$ is chosen as described in \cite{Jaime:2012gc}. 
The   distance modulus given by
\be
\mu=m-M = 5\log_{10}\frac{d_L^{flat}}{Mpc}+25 
\ee
is the quantity reported by supernovae data.
\begin{figure}[h!]
 \centering
 \includegraphics[width=4.8cm,height =7.6cm,angle = -90]{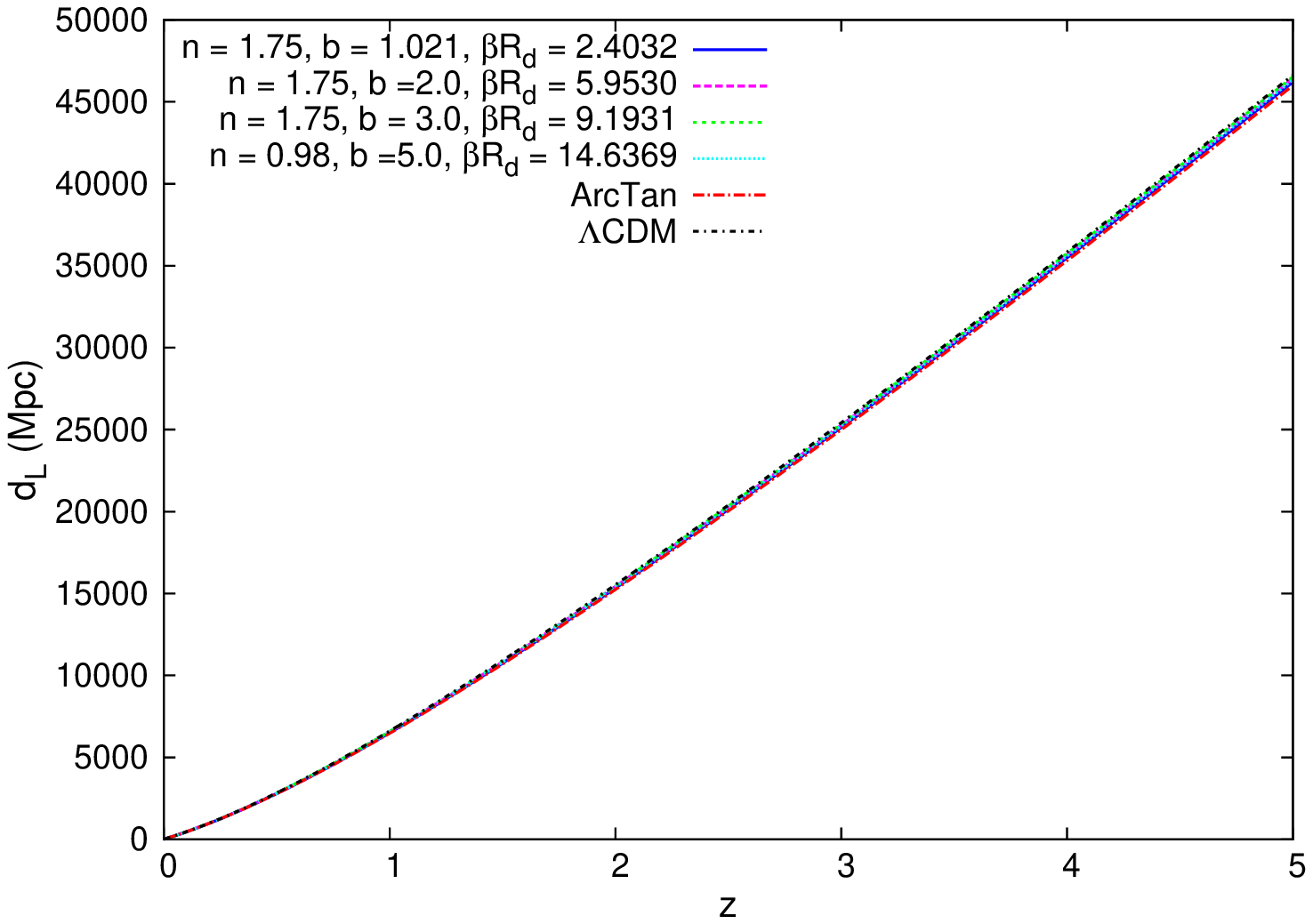}
 \caption{Luminosity distance vs $z$}
 \label{fig:dlvsz}
\end{figure}
\begin{figure}[h!]
 \centering
 \includegraphics[width=7.6cm,height =4.8cm,angle = 0]{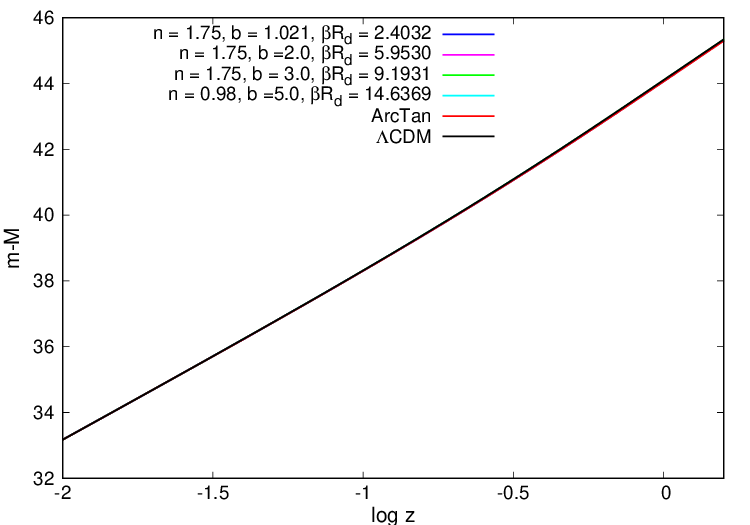}
 \caption{Distance modulus vs $z$ }
 \label{fig:dmvsz}
\end{figure}
Figs.~\ref{fig:dlvsz} and \ref{fig:dmvsz} show variation of luminosity distance and distance modulus w.r.t $z$ for 
the $f(R)$ model (\ref{model}) along with the $\Lambda$CDM. 
The luminosity distance and distance modulus for the model are not much different from 
the $\Lambda$CDM for all choice of parameters. Distance modulus is also
plotted along with the supernovae data of  UNION 2 \cite{Amanullah:2010vv} in Figs.~\ref{fig:spnvdata1} and \ref{fig:spnvdata2}. In the next section, we apply $Omh^2$ diagnostic to test $f(R)$ model written in \eqrf{model}. 


\section{$Omh^2$ Diagnostic}\label{sec:omh2}
Starting with the Hubble parameter, Sahni et al.\cite{Sahni:2008xx} proposed a new diagnostic namely $Om$ diagnostic:
\be
Om(z)=\frac{\bar h^2(z)-1}{(1+z)^3-1},
\label{om1}
\ee
where $\bar h(z)=H(z)/H_0$, with $H_0$ being value of Hubble parameter today. It is noticeable that the function $Om(z)$ remains constant at $\Omega_M^0$ for all $z$ in $\Lambda$CDM model, where $\Omega_M^0$ is the present density parameter. While in other evolving Dark Energy models and also in modified gravity theories $Om(z)$ varies with redshift $z$, its value remains constant at $\Omega_M^0$ in $\Lambda$CDM model. Thus, the deviation from the relation $Om(z)-\Omega_M^0=0$ would indicate that the late-time acceleration is not due to cosmological constant only. If $Om(z)$ is plotted against $(1+z)^3$ then the plot is a horizontal line for $\Lambda$CDM but not in modified gravity and thus any small departure in the behavior of modified gravity from $\Lambda$CDM can be identified. One remarkable fact about this diagnostic is that it depends only on the Hubble parameter and therefore it is easy to determine its values from observations. One can employ observed values of $H(z)$ from different observations to construct $Om(z)$ in the statistically independent way. The function $Om(z)$ written in \eqrf{om1} can also be written as two-point diagnostic $Om(z,0)$. Generalizing this we have a two-point diagnostic
\be
Om(z_i;z_j)=\frac{\bar h^2(z_i)-\bar h^2(z_j)}{(1+z_i)^3-(1+z_j)^3}
\label{om2}
\ee
If the value of Hubble parameter is known at two or more redshifts then constructing the $Om(z_i;z_j)$ one can test $\Lambda$CDM as well as any modified gravity theory as a model explaining the late-time acceleration. 

Another paper, Sahni et al. \cite{Sahni:2014ooa}, gave an improved diagnostic multiplying by $h^2=H_0^2/100^2$ both sides of \eqrf{om2},
\be
Omh^2(z_i;z_j)=\frac{h^2(z_i)-h^2(z_j)}{(1+z_i)^3-(1+z_j)^3},
\label{om3}
\ee
where $h(z)=H(z)/100$.
Since $Omh^2=\Omega_M^0h^2$ for $\Lambda$CDM and the value of $\Omega_M^0h^2$ is provided by cosmic microwave background observations this "improvement" in $Omh^2$ diagnostic gives the leverage to test the $\Lambda$CDM model easily. Using Planck XVI 2013 results \cite{Ade:2013zuv}, $\Omega_M^0h^2=0.1426\pm 0.0025$, and measurements of Hubble parameter from Baryon Acoustic Oscillation observations in Sloan Digital Sky Survey \cite{Samushia:2012iq,Anderson,Delubac:2014aqe}, it is found that $Omh^2$ indeed varies with redshift $z$ and gives departure from $Omh^2=\Omega_M^0h^2$ showing tension with $\Lambda$CDM at $2\sigma$ \cite{Sahni:2014ooa}. They considered three different redshifts, $z_1=0$, $z_2=0.57$ and $z_3=2.34$ and corresponding Hubble parameters $H(z_1)=70.6\pm3.3\;km\;s^{-1}Mpc^{-1}$ \cite{Efstathiou}, $H(z_2)=92.4\pm4.5\;km\;s^{-1}Mpc^{-1}$ \cite{Samushia:2012iq} and $H(z_3)=222\pm7\;km\;s^{-1}Mpc^{-1}$ ‪\cite{Delubac:2014aqe}. The values of two-point relations $Omh^2$ reported by \cite{Sahni:2014ooa} are
\bea
Omh^2(z_1;z_2)=0.124\pm0.045,\nonumber \\
Omh^2(z_1;z_3)=0.122\pm0.010,\nonumber \\
Omh^2(z_2;z_3)=0.122\pm0.012.
\label{om4}
\eea
In \cite{Sahni:2014ooa}, it is shown that these values are not sensitive to the values of $H_0$. Also, values of Hubble parameter at redshifts $z_1,z_2$ and $z_3$ are obtained from three different observations and therefore $Omh^2$ given in \eqrf{om4} are model independent. We apply this model-independent diagnostic to test our model in next section.

\subsection{Diagnostic of $f(R)$ Model}\label{ssec:diagnosticofmodel}
\begin{figure*}
  \centering
   $
   \begin{array}{c c}
   \includegraphics[width=0.5\textwidth]{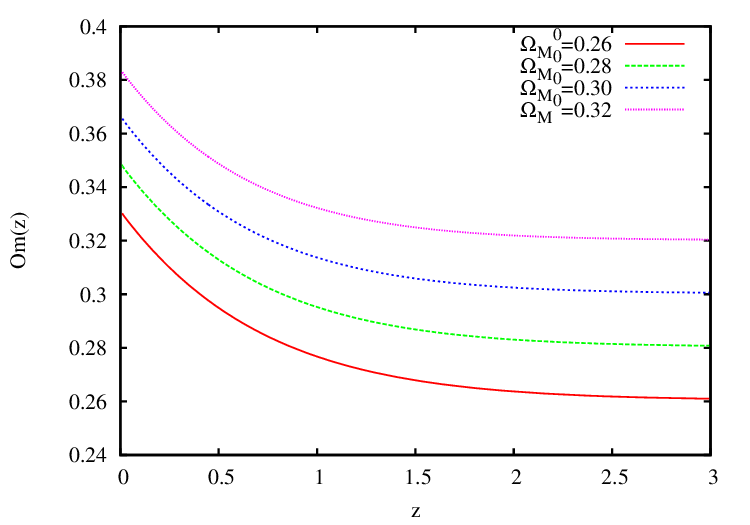} & 
   \includegraphics[width=0.5\textwidth]{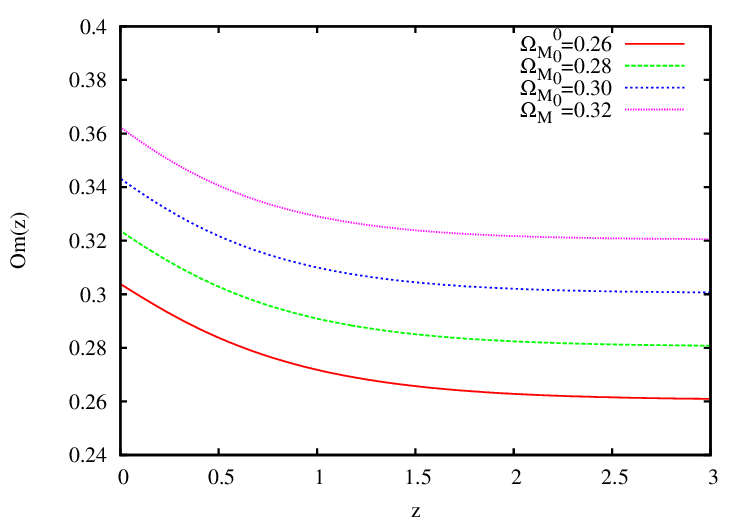}
   \end{array}
   $ 
  \caption{$Om(z)$ vs $z$ for $b=1.021$ and $n=1.75$ in left and $b=1.5$ and $n=1.75$ in right.  }
  \label{fig:omz1}
\end{figure*}

\begin{figure*}
  \centering
   $
   \begin{array}{c c}
   \includegraphics[width=0.5\textwidth]{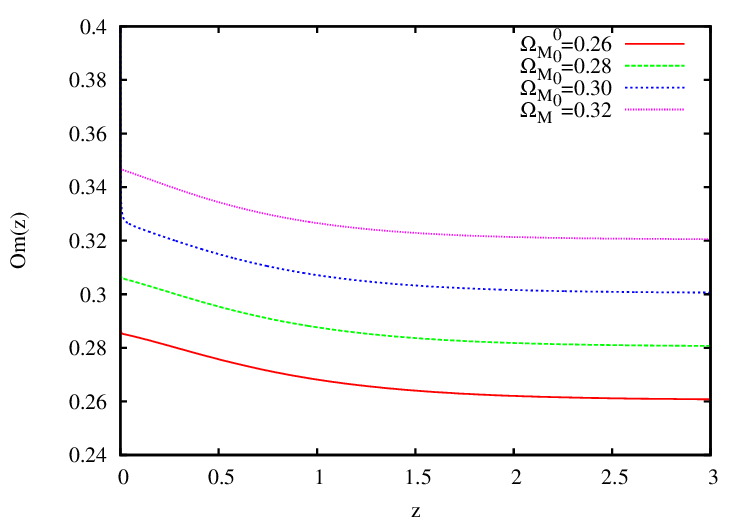} & 
   \includegraphics[width=0.5\textwidth]{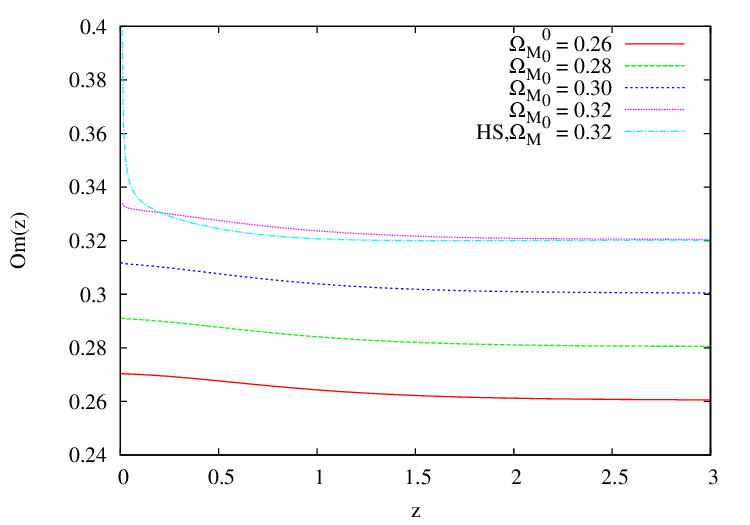}
   \end{array}
   $ 
  \caption{$Om(z)$ vs $z$ for $b=2.0$ and $n=1.75$ in left and $b=3.0$ and $n=1.75$ in right. The curve labeled by 'HS' is Hu-Sawicki model }
  \label{fig:omz2}
\end{figure*}

\begin{figure*}
  \centering
   $
   \begin{array}{c c}
   \includegraphics[width=0.5\textwidth]{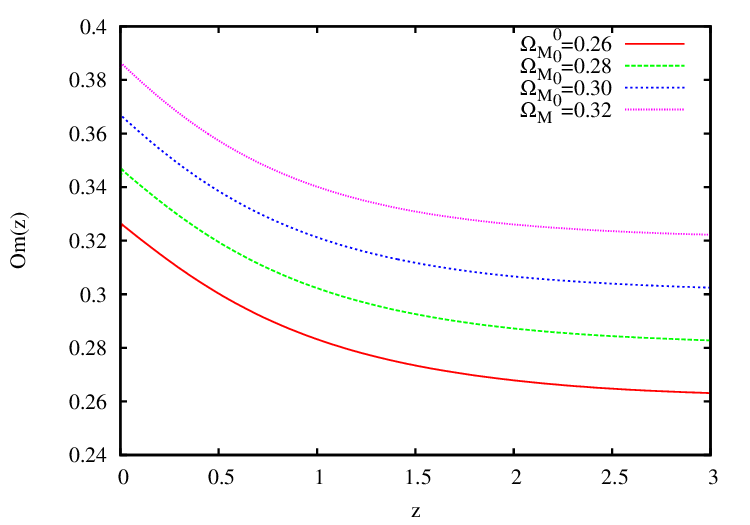} & 
   \includegraphics[width=0.5\textwidth]{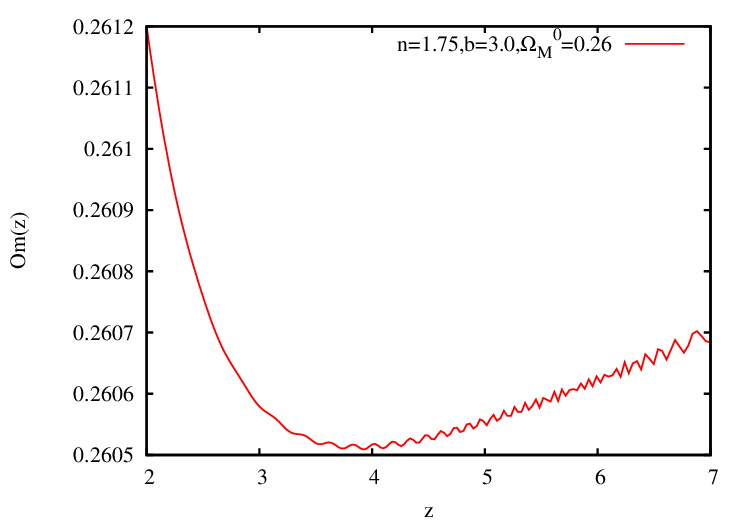}
   \end{array}
   $ 
  \caption{$Om(z)$ vs $z$ for $b=2.0$ and $n=1.0$ in left and $b=3.0$ and $n=1.75$ and $\Omega_M^0=0.26$ in right.  }
  \label{fig:omz3}
\end{figure*}

Through this Om diagnostic, we can detect the distinguishability of $f(R)$ model from $\Lambda$CDM. In addition to this, we can pull the best choice of parameter values which fit well with the observations by calculating $\chi^2$ values of two-point relations $Omh^2$. To acquire the theoretical $Omh^2(z_i;z_j)$, one needs to solve the cosmological evolution equations and fetch the values of $H$ at different $z$ for a given model. We establish the cosmological evolution equations \eqref{deq1}, \eqref{deq2}, \eqref{deq3} and \eqref{deq4} and solve them in the preceding section. Plugging the calculated values of $H(z)$ in \eqrf{om1}, we obtain the theoretical values of $Om(z)$. Here, we are interested in comparing $Om(z)$ profile for different $\Omega_M^0$ and therefore we use different values of $\Omega_M^0$ to solve the cosmological equations. In Figs.~[\ref{fig:omz1}], [\ref{fig:omz2}], and left of Fig.~[\ref{fig:omz3}], $Om(z)$ functions are plotted w.r.t. redshift $z$ for different $\Omega_M^0$ values for different parameters listed in Table \ref{tabl:desitter}.  
It can be easily seen that value of $Om(z)$ remains constant around $z=3$ but at lower redshifts $z<2$ the $Om(z)$ increases as $z$ decreases. This clearly indicates the deviation from $\Lambda$CDM and implies that the universe explained by the model given in \eqrf{model} behaves as $\Lambda$CDM for higher redshifts but deviates for $z<2$ i.e. in late time epoch. The deviation grows more for lower $n$ and $b$ being maximum for $n=1.75$ and $b=1.021$. We can also see that the typical feature of $f(R)$ gravity first noticed for Starobinsky and Hu-Sawicki models in \cite{Jaime:2015afa} is also present in our model which is shown for a particular parameter value in right figure of Fig.~[\ref{fig:omz3}]. 

To construct a two-point function $Omh^2(z_i;z_j)$, we consider three redshifts $z_1=0$, $z_2=0.57$ and $z_3=2.34$ for which observational $H^2(z)$ values are known from three independent experiments. For these three redshifts, observational values of $Omh^2$ are written in \eqrf{om4}. The parameter value for best fit $Omh^2$ with observations can be found by computing $\chi^2$.
\begin{table*}
\centering
\begin{tabular}{|l|l|l|l|l|l|l|r|}
  \hline
      b & n   & $\Omega_M^0=0.26$  & $\Omega_M^0=0.28$ & $\Omega_M^0=0.30$  &  $\Omega_M^0=0.32$  &  $\Omega_M^0=0.34$ \\
  \hline
  $1.021$ & $1.75$    &  0.7943  &  4.5159  &  11.4596  &  21.6450  &  35.0983    \\
  $1.1$   & $1.75$    &  0.7769  &  4.5003  &  11.4547  &  21.6625  &  35.1382    \\
  $1.5$   & $1.75$    &  0.7002  &  4.4105  &  11.3895  &  21.6492  &  35.2001    \\
  $2.0$   & $1.75$    &  0.6356  &  4.3092  &  11.2816  &  21.5600  &  35.1484    \\  
  $3.0$   & $1.75$    &  0.5772  &  4.1967  &  11.1414  &  21.4138  &  35.0176    \\  
  $5.0$   & $0.98$    &  0.7751  &  4.6681  &  11.8557  &  22.3316  &  36.0988    \\
  $2.0$   & $1.0$     &  1.2121  &  5.4699  &  12.9336  &  23.5972  &  37.4733    \\
  \hline
\end{tabular}
\caption{$\chi^2$ values for different $b$ and $n$ and different values of $\Omega_M^0$}
\label{tabl:chisquare}
\end{table*}
From Table \ref{tabl:chisquare}, it is evident that the best-fit parameters are $b=3.0$ and $n=1.75$ because $\chi^2$ is minimum for this choice of parameters. For $\Lambda$CDM, $\chi^2$ comes out to be $7.361$ for $\Omega_M^0=0.28$ which is much higher than $\chi^2$ values written in Table \ref{tabl:chisquare}. Thus, from the simple observation based $Omh^2$ diagnostic shows that the $f(R)$ model given in \eqrf{model} is more prefered than $\Lambda$CDM in $z\leq 2$ regime. By comparing the $\chi^2$ values obtained in \cite{Jaime:2015afa} with the values given in above Table \ref{tabl:chisquare}, one can say that the model written in \eqrf{model} fits better with observations than the models considered in \cite{Jaime:2015afa}.

\section{Conclusions}\label{sec:conclusion}
To cure the incompetence of the Arctan model proposed in \cite{Kruglov:2013qaa} to evade the local gravity tests \cite{Dutta:2016ukw}, we extend the model to a new model given in \eqrf{model}. Here, we show that our model can pass local gravity tests, constraining the model parameters for the fifth-force condition to be held.

The $f(R)$ model written in \eqref{model} can mimic as an effective cosmological constant plus GR when the curvature of the universe is very large. As depicted in Fig.~\ref{fig:functionnewarctan}, taking large value of power $n$ can make the function $F(R)$ increases more rapidly to a constant value therefore the proper range of $n$ should be $1< n\leq2$. We also find that the stable fixed points of the model are $P_5$ and $P_1$ exhibiting the evolution of the universe from saddle matter era to late-time acceleration. 

We put the model in \eqref{model} under scrutiny to investigate the fifth-force constraints for the model through chameleon mechanism. We find that the model can evade local gravity test for the parameter values $n\geq 1.75$ and $b\geq 1.021$. These constraints on the parameter values give rise to constraint on the value of corresponding de Sitter point which is $\beta R_d\geq 2.4$. 

The existence of curvature singularity in the model in \eqrf{model} is also investigated. We find that there may be finite probability for scalar field to hit the singularity due to the finite potential $V_J$ at the singular point during its evolution. As described in \cite{Appleby:2009uf}, the curvature singularity can be cured by adding $R^2$ term to the Lagrangian.

We also derive the cosmological evolution equations for this model considering an FRW background universe and solve them numerically. We obtain the oscillatory behavior of Ricci scalar $R$ and Hubble parameter $H$ which becomes less pronounced with higher values of $b$. It is found that the model \eqref{model} provides sufficiently long matter dominated era and radiation era at larger redshifts similar to $\Lambda$CDM model. These results are reconciled with the predictions of fixed point analysis and nucleosynthesis. The equation of state of the $f(R)$ component of the fluid $w_x$ for our model oscillates around phantom divide $w=-1$. We also find that the luminosity distance and distance modulus obtained in our model are not much different from $\Lambda$CDM model and also fit with the supernova data. On the other hand, through $Om$ diagnostic we find that the model given in \eqrf{model} is distinguishable from $\Lambda$CDM at redshifts $z<2$. In $Om$ diagnostic analysis, our model emerges as a better fit with baryon acoustic oscillation data than $\Lambda$CDM. Performing $\chi^2$ test, we obtain $n=1.75$ and $b=3.0$ as a best-fit choice of parameters. 

The distinguishability from $\Lambda$CDM and also the viability of any $f(R)$ model can be better tested with large-scale structure data. The recent redshift space distortion(RSD) data can put tighter constraints on the model, but yet an available number of data points is very less to test viability of any model in a definite manner. Thus, we conclude that the model given in \eqrf{model} gives rise to the dark energy dominated era while having correct cosmic history to coincide with other cosmological observations and also evade local gravity tests.  

%
%

\end{document}